\documentclass[useAMS]{mn2e}
\usepackage{times}
\usepackage{graphicx}

\title{{\it Chandra\/} observations of five ultraluminous X-ray sources in
nearby galaxies}

\author[T.\,P. Roberts et al.]
{T.\,P. Roberts$^{1,*}$, R.\,S. Warwick$^1$, M.\,J. Ward$^1$ \&
M.\,R. Goad$^2$\\ $^1$ X-ray and Observational Astronomy Group,
Dept. of Physics \& Astronomy, University of Leicester, University
Road, Leicester, LE1 7RH\\ $^2$ Dept. of Physics \& Astronomy,
University of Southampton, Highfield, Southampton, Hants., SO17 1BJ\\
$^*$E-mail: tro@star.le.ac.uk}
\date{}

\def\ro{{\it ROSAT~\/}}
\def\asca{{\it ASCA~\/}}
\def\ein{{\it Einstein~\/}}
\def\xmm{{\it XMM-Newton~\/}}

\def\hst{{\it HST~\/}}
\def\chan{{\it Chandra~\/}}

\def\ergcms{{\rm ~erg~cm^{-2}~s^{-1}}}

\def\ergsec{{\rm ~erg~s^{-1}}}

\def\atpcm{{\rm ~atoms~cm^{-2}}}
\def\ctsec{{\rm ~count~s^{-1}}}

\def\H0{{\rm ~km~s^{-1}~Mpc^{-1}}}

\def\kmsec{{\rm ~km~s^{-1}}}

\def\la{\mathrel{\hbox{\rlap{\hbox{\lower4pt\hbox{$\sim$}}}{\raise2pt\hbox{$<$}}}}}
\def\ga{\mathrel{\hbox{\rlap{\hbox{\lower4pt\hbox{$\sim$}}}{\raise2pt\hbox{$>$}}}}}

\def\d25{D$_{25}$}
\def\nh{{$N_{H}$}}

\def\hi{H{\small I}$~$}
\def\hii{H{\small II}$~$}

\def\.25{0.25 keV\thinspace}
\def\lx{L$_{\rm X}$}

\begin{document}

\maketitle

\begin{abstract}
We report the results of a programme of dual-epoch \chan ACIS-S
observations of five ultraluminous X-ray sources (ULXs) in nearby
spiral galaxies.  All five ULXs are detected as unresolved, point-like
X-ray sources by {\it Chandra\/}, though two have faded below the $10^{39}
\ergsec$ luminosity threshold used to first designate these sources as 
ULXs.  Using this same criterion, we detect three further ULXs within
the imaged regions of the galaxies.  The ULXs appear to be related to
the star forming regions of the galaxies, indicating that even in
``normal'' spiral galaxies the ULX population is predominantly
associated with young stellar populations.  A detailed study of the
\chan ACIS-S spectra of six of the ULXs shows that five are better
described by a powerlaw continuum than a multi-colour disc blackbody
model, though there is evidence for additional very soft components to
two of the powerlaw continua.  The measured photon indices in four out
of five cases are consistent with the low/hard state in black hole
binaries, contrary to the suggestion that powerlaw-dominated spectra
of ULXs originate in the very high state.  A simple interpretation of
this is that we are observing accretion onto intermediate-mass black
holes, though we might also be observing a spectral state unique to
very high mass accretion rates in stellar-mass black hole systems.
Short-term flux variability is only detected in one of two epochs for
two of the ULXs, with the lack of this characteristic arguing that the
X-ray emission of this sample of ULXs is not dominated by
relativistically-beamed jets.  The observational characteristics of
this small sample suggest that ULXs are a distinctly heterogeneous
source class.
\end{abstract}


\section{Introduction}

Ultraluminous X-ray Sources (ULXs) can be broadly defined as the most
luminous point-like extra-nuclear X-ray sources located within nearby
galaxies, displaying X-ray luminosities in excess of $10^{39}
\ergsec$.  These sources were first observed in \ein observations of
nearby galaxies (e.g. Fabbiano \& Trinchieri 1987), and more than one
hundred were subsequently catalogued in \ro high-resolution imager
observations (Roberts \& Warwick 2000; Colbert \& Ptak 2002).  Whilst
a fraction of this observed ULX population is associated with recent
supernovae (e.g. SN 1986J, Bregman \& Pildis 1992; SN 1979C, Immler,
Pietsch \& Aschenbach 1998), \asca studies have shown that many ULXs
appear to display the characteristics of accreting black holes
(e.g. Makishima et al. 2000; Mizuno, Kubota \& Makishima 2001).
Crucially, high spatial resolution \chan observations (e.g. Kaaret et
al. 2001; Strickland et al. 2001), have failed to resolve most ULX
targets.  When combined with the observed significant flux
variability, this suggests the presence of a single, luminous source
of X-rays rather than a grouping of less luminous sources.

However, debate over the nature of the bulk of the ULX population is
ongoing, since if these are accretion-powered sources their X-ray
luminosities match, and in many cases greatly exceed, the Eddington
limit for a typical stellar-mass ($\sim 10$ M$_{\odot}$) black hole.
Suggestions for their physical composition currently focus upon four
possibilities.  The first is that many ULXs are accreting examples of
a new class of $10^2 - 10^5$ M$_{\odot}$ intermediate-mass black holes
(IMBH; e.g. Colbert \& Mushotzky 1999).  The formation of such objects
remains a matter of some debate; some intermediate-mass black holes
may be the remnants of primordial Population III stars (e.g. Madau \&
Rees 2001), whilst others may be formed in dense globular clusters
prior to being deposited in a galaxy disc (Miller \& Hamilton 2002).
An alternative scenario is that IMBH are formed by the runaway merger
of stellar objects at the centre of young, dense stellar clusters
(Ebisuzaki et al. 2001; Portegies Zwart \& McMillan 2002).  The
existence of IMBH may be supported by the recent inferrence of $3
\times 10^3$ and $2 \times 10^4$ M$_{\odot}$ massive dark objects in
the cores of the globular clusters M15 and G1 respectively (van der
Marel et al. 2002; Gerssen et al. 2002; Gebhardt et al. 2002), though
the presence of such an object in M15 is far from proven (Baumgardt et
al. 2002).  Further evidence has recently emerged with the discovery
of the X-ray spectral signature of ``cool'' accretion discs,
consistent with the presence of IMBH, in several ULXs (e.g. Miller et
al. 2003, Roberts \& Colbert 2003).

The remaining models focus upon interpreting ULXs as extreme examples
of ``ordinary'' stellar-mass (i.e. $\sim 10$ M$_{\odot}$) black hole
X-ray binaries.  The second physical model is that many ULXs are
ordinary X-ray binaries in an unusually high accretion mode, in which
their accretion disc becomes radiation pressure-dominated, producing
photon bubble instabilities that allow the disc to radiate at a truly
super-Eddington X-ray flux (Begelman 2002).  Thirdly, many ULXs may
only appear to exceed the Eddington limit, but could in fact be X-ray
binaries emitting anisotropically (King et al. 2001), with only a mild
beaming factor $b \sim 0.1$ (where $b = \Omega/4\pi$, and $\Omega$ is
the solid angle of the X-ray emission) required in most cases to
reduce the energy requirements below the Eddington limit for a
conventional stellar-mass black hole.  The fourth model is a variation
on the third scenario, suggested by Reynolds et al. (1997) and more
recently K{\"o}rding, Falcke \& Markoff (2001) and Georganopoulos,
Aharonian \& Kirk (2002), in which ULXs are microquasars in nearby
galaxies that we are observing directly down the beam of their
relativistic jet (``microblazars'', c.f. Mirabel \& Rodriguez 1999).
Observational support for this last scenario comes from the detection
of radio emission, potentially the signature of a
relativistically-beamed jet, emanating from an ULX in NGC 5408 (Kaaret
et al. 2003).

Current observations do not completely rule out any of the above
scenarios.  However, they do suggest that there may be at least two
separate underlying populations of ULXs, since a large number are seen
coincident with active star formation regions (e.g. Zezas \& Fabbiano
2002; Roberts et al. 2002) and hence are presumably associated with
nascent stellar populations, whereas some ULXs are found in elliptical
galaxies (e.g. Irwin, Athey \& Bregman 2003) and so must be associated
with an older stellar population.  King (2002) suggests that the
population associated with star formation are high-mass X-ray binaries
(HMXBs) undergoing an episode of thermal-timescale mass transfer (see
also King et al. 2001), whilst the older population may be
long-lasting transient outbursts in low-mass X-ray binaries.  Local
examples of each suggested class are SS 433 and GRS 1915+105
respectively.  This heterogeneity is supported by the first reported
optical stellar counterparts to ULXs.  Roberts et al. (2001) report
the detection of a blue continuum source coincident with NGC 5204 X-1,
which \hst resolved into three separate sources with colours that are
consistent with young, compact stellar clusters in NGC 5204 (Goad et
al. 2002).  \hst observations also show an ULX in M81 (NGC 3031 X-11)
to have a possible O-star counterpart (Liu et al. 2002).  In contrast,
ULXs have been found with potential globular cluster counterparts in
both NGC 4565 (Wu et al. 2002) and NGC 1399 (Angelini, Loewenstein \&
Mushotzky 2001), suggesting that these ULXs are associated with the
older stellar population, or perhaps massive central black holes, of
these globular clusters.

In this paper, we present the results of dual-epoch \chan observations
of five different ULXs located in nearby ($d < 10$ Mpc) galaxies,
awarded \chan time in AO-2 \& AO-3.  These ULXs are listed in
Table~\ref{obsdetails}.  They were selected from the catalogues of \ro
HRI point-like X-ray source detections in nearby galaxies presented by
Roberts \& Warwick (2000; hereafter RW2000) and Lira, Lawrence \&
Johnson (2000) on the basis of their high X-ray luminosities (L$_{\rm
X} > 10^{39} \ergsec$ in the \ro HRI) and the lack of previous \chan
observations.  A discussion of the observational history of these ULXs
is presented in Appendix A.  Further discussion of the environment of
each of these ULXs, as determined from William Herschel
Telescope/INTEGRAL IFU observations, will be detailed in a future
paper (Roberts et al., in prep.).

The layout of this paper is as follows.  In the next section we
briefly outline the details of the \chan observations and the data
reduction.  This is followed by a discussion of the detection of these
(and three other) ULXs, and their locations within their host
galaxies.  In section 4 we detail the spatial, temporal and spectral
properties of each source, before discussing the implications of these
results for possible physical models of ULXs in section 5.  Our
findings and conclusions are summarised in section 6.

\begin{table*}
\caption{Details of the ten \chan observations.}
\begin{tabular}{lcccccc}\hline
Target	& \multicolumn{2}{c}{Observation aimpoint}  & \chan sequence
number	 & Observation date	& ACIS-S3 subarray	& Exposure \\
 & Right ascension	& Declination	& & (yyyy-mm-dd)	& &
(ks) \\\hline 
IC 342 X-1	& $03^h45^m55.2^s$	& $+68^{\circ}04'55''$	& 600253 
& 2002-04-29	& ${1}\over{8}$	& 9.9 \\
& & & 600254 & 2002-08-26	& & 9.9 \\
NGC 3628 X-2	& $11^h20^m37.5^s$	& $+13^{\circ}34'28''$	& 600255 
& 2002-04-06	& {\it full\/}	& 22.3 \\
& & & 600256 & 2002-07-04	& & 22.5 \\
NGC 4136 X-1	& $12^h09^m22.6^s$	& $+29^{\circ}55'49''$	& 600257 
& 2002-03-07	& {\it full\/}	& 18.8 \\
& & & 600258 & 2002-06-08	& & 19.7 \\
NGC 4559 X-1	& $12^h35^m52.0^s$	& $+27^{\circ}56'01''$	& 600160 
& 2001-01-14	& ${1}\over{4}$	& 9.7 \\
& & & 600161 & 2001-06-04	& & 11.1 \\
NGC 5204 X-1	& $13^h29^m39.2^s$	& $+58^{\circ}25'01''$	& 600162 
& 2001-01-09	& ${1}\over{8}$	& 10.1 \\
& & & 600163 & 2001-05-02	& & 9.5 \\\hline
\end{tabular}
\label{obsdetails}
\end{table*}

\section{Chandra observations and data reduction}

The details of the ten \chan observations of ULXs that form the basis
of this paper are listed in Table~\ref{obsdetails}.  The observations
were pointed at the \ro HRI positions of the ULXs, after a correction
to the astrometry calculated from X-ray/optical matches in the HRI
field-of-view.  All coordinates listed in the Table, and throughout
this paper, are epoch J2000.  We tabulate some basic parameters for
the host galaxies in Table~\ref{galparams}.  The observations were
performed between 2001 January 9 and 2002 August 26, and range in
exposure time from 9.7 to 22.5 ks.  Each target was observed on two
occasions, separated by 3 -- 5 months.  In order to mitigate the
anticipated effects of detector pile-up in the ACIS-S S3 chip, several
of the observations were performed in a sub-array mode; these are
listed in Table~\ref{obsdetails}.  The choice of sub-array was
governed by the anticipated count rates, modelled from the previous
\ro HRI observations of these objects (RW2000; Lira, Lawrence
\& Johnson 2000), with the aim of limiting the pile-up fraction to
10\% or less in each observation.  On this basis, sub-arrays were
deemed necessary for three of the five targets in the programme.

\begin{table}
\caption{The host galaxies.}
\begin{tabular}{lcccc}\hline
Galaxy	& Hubble type$^a$	& $d ^b$& $i ^b$ & \nh $^c$ \\
	&		& (Mpc)	& ($^{\circ}$)	&  ($\times 10^{20}$
cm$^{-2}$) \\\hline 
IC 342	& SAB(rs)cd	& 3.9	& 20	& 30.3 \\
NGC 3628& SAb pec sp	& 7.7	& 87	& 2.0 \\
NGC 4136& SAB(r)c	& 9.7	& 0$^d$	& 1.6 \\
NGC 4559& SAB(rs)cd	& 9.7	& 69	& 1.5 \\
NGC 5204& SA(s)m	& 4.8	& 53	& 1.5 \\\hline
\end{tabular}
Notes:\\ $^a$ Data from the NASA/IPAC extragalactic database (NED).\\
$^b$ Distance ($d$) and inclination ($i$) data from Tully (1988).\\
$^c$ Foreground absorption, interpolated at the position of each
galaxy from the \hi maps of Stark et al. (1992).\\ $^d$ There is no
recorded inclination, so a face-on aspect is adopted after inspection
of Digitised Sky Survey images.
\label{galparams}
\end{table}

Data reduction was performed using the {\small CIAO} software suite,
versions 2.1 and 2.2.  The reduction started in each case with the
level two event file, from which events with energies outside the 0.3
-- 10 keV range were rejected.  All datasets were searched for periods
of high background flaring, but none were observed, allowing the full
science exposure to be utilised in each case.  Further steps in the
analysis of the data are outlined in the following sections.

\section{ULX detections and locations}

A 0.3 -- 10 keV image of the central $8.1' \times 8.1'$ region of each
field was constructed from the corresponding cleaned event file.  This
image was searched for point sources using {\small WAVDETECT}, a
wavelet-based source detection algorithm available in the {\small
CIAO} package, and the X-ray source detections coincident with the
optical extent of each galaxy (or that part of it within the ACIS-S3
field-of-view) were catalogued.  The target ULXs were detected in
every observation, and their \chan nomenclature\footnote{We refer to
the ULXs by their \chan names throughout this paper.  The
cross-identification with previous names is shown where necessary.},
including their refined position, and observed count rates are listed
in Table~\ref{ulxs}.  Count rates were converted to approximate fluxes
based on a simple powerlaw continuum model with a standard photon
index ($\Gamma$) of 2, subject to a foreground absorption column
appropriate for each galaxy (see Table~\ref{galparams}).  In addition
to the targets, three other sources with luminosities potentially in
the ULX regime were found; one is a previously catalogued source (NGC
4559 X-4 in RW2000), but the other two are new ULX identifications.
The details of these new ULXs are also given in Table~\ref{ulxs}.

\begin{table*}
\caption{The ULX detections.}
\begin{tabular}{lcrrcccl}\hline
CXOU J	& Previous	& \multicolumn{2}{c}{Count rate (ct
ks$^{-1}$)}	& \multicolumn{3}{c}{Offset from galaxy nucleus}
& Location \\
	& designation	& Epoch 1	& Epoch 2 &  observed ($''$)
& deprojected (kpc)	& $f$(R$_{25}$)	& \\\hline
{\it Targetted ULX\/} \\
034555.7+680455	& IC 342 X-1	& $212 \pm 5$	& $226 \pm 5$	& 302
& 6.1   & 0.68  & Spiral arm \\
112037.3+133429	& NGC 3628 X-2	& $13 \pm 1$	& $13 \pm 1$	& 292
& 10.8  & 0.66  & Outer disc \\
120922.6+295551	& NGC 4136 X-1	& $3 \pm 1$	& $4 \pm 1$ 	& 65
& 3.0   & 0.54  & Spiral arm\\
123551.7+275604	& NGC 4559 X-1	& $153 \pm 4$	& $192 \pm 4$ 	& 123
& 15.2  & 0.96  & Faint outer spiral arm\\
132938.6+582506	& NGC 5204 X-1	& $411 \pm 6$	& $159 \pm 4$ 	& 17
& 0.7   & 0.21  & Inner disc \\
{\it Field ULX\/} \\
120922.2+295600	& -	& $26 \pm 1$	& $18 \pm 1$ 	& 62	& 2.9
& 0.52  & Spiral arm \\
123557.8+275807	& -	& $11 \pm 1$	& $21 \pm 1$ 	& 31	& 2.3
& 0.15  & Inner disc \\
123558.6+275742	& NGC 4559 X-4	& $62 \pm 3$	& $ 119 \pm 3$ 	& 12
& 1.7	& 0.11	& Inner disc/bulge \\\hline
\end{tabular}
\label{ulxs}
\end{table*}

One excellent capability of the \chan observatory is that it provides
an absolute astrometry solution to an accuracy of one arcsecond or
better.  This positional accuracy facilitates detailed follow-up of
the \chan X-ray source detections through the characterisation of
their multi-wavelength counterparts.  This avenue has already borne
fruit in the study of ULXs, as discussed in the introduction.
However, here we limit ourselves to a brief discussion of the
environment of the \chan ULX detections on the basis of Palomar
Digitised Sky Survey (DSS) data.

The positions of the ULXs relative to their host galaxies are
illustrated in Figures~\ref{ulxlocations} \& \ref{moreulxlocations},
where we overlay the \chan X-ray emission contours onto DSS-2 (blue)
images of the equivalent field-of-view.  In each case the target ULX
is at the centre of the field-of-view.  The position of each ULX with
respect to the nucleus of the parent galaxy is quantified in
Table~\ref{ulxs}.  The observed offset from the position of the galaxy
nucleus, taken in each case from Falco et al. (1999), is simply the
projected distance measured in arcseconds; we correct this to a
deprojected radius, assuming that each source is in the plane of its
host galaxy, using the host's inclination as given in
Table~\ref{galparams} and a position angle taken from the RC3
catalogue (de Vaucouleurs et al. 1991).  This deprojected radius is
also shown as $f$(R$_{25}$), the fraction of the distance between the
nucleus and the edge of the galaxy, defined by the semi-major axis of
the 25 mag arcsec$^{-2}$ isophotal ellipse, at which the ULX is found.
Finally, we also give a qualitative description of the region of the
galaxy in which the ULX is found.

\begin{figure*}
\centering
\includegraphics[width=8cm,]{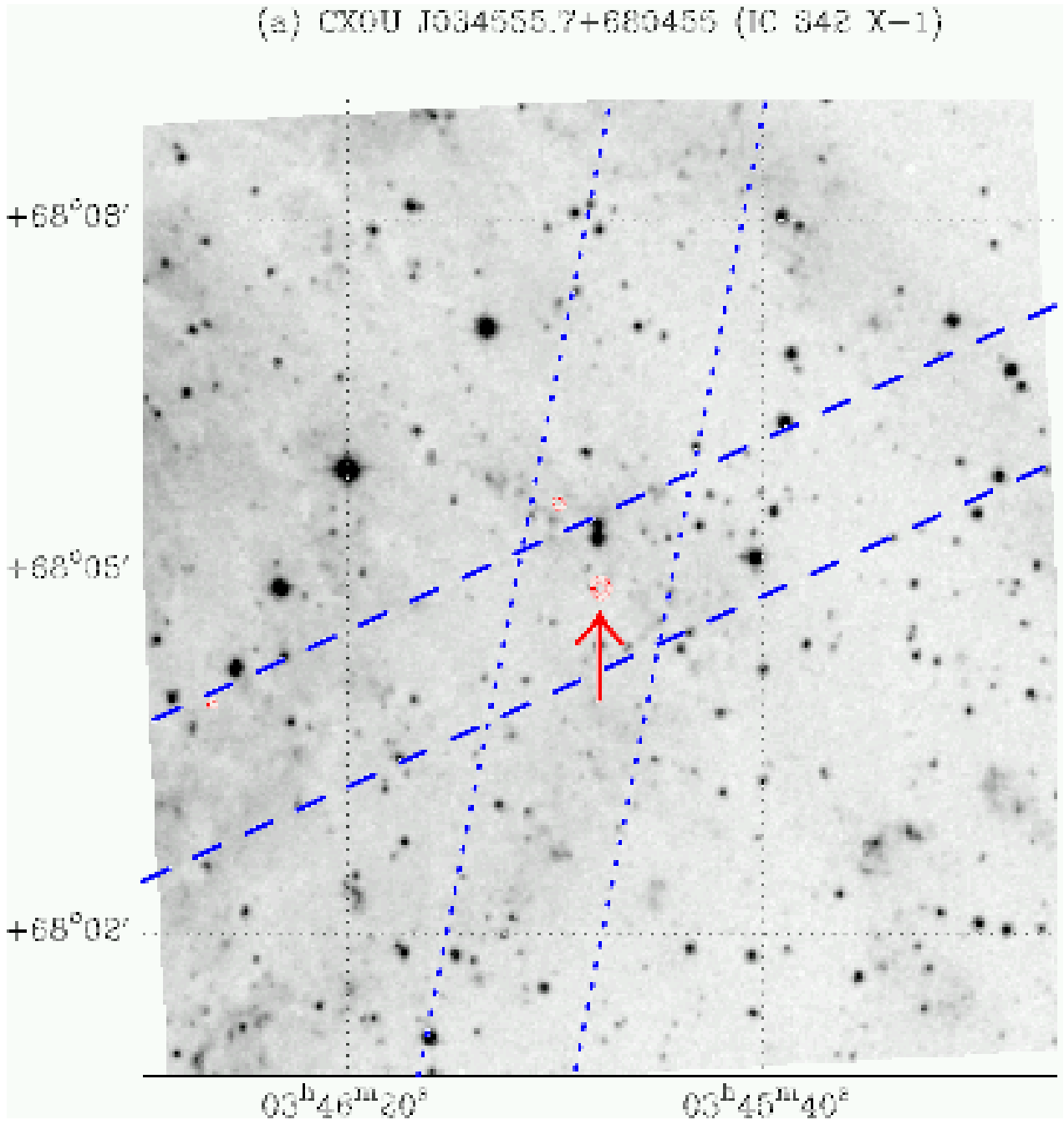}
\hspace*{2mm}
\includegraphics[width=8cm]{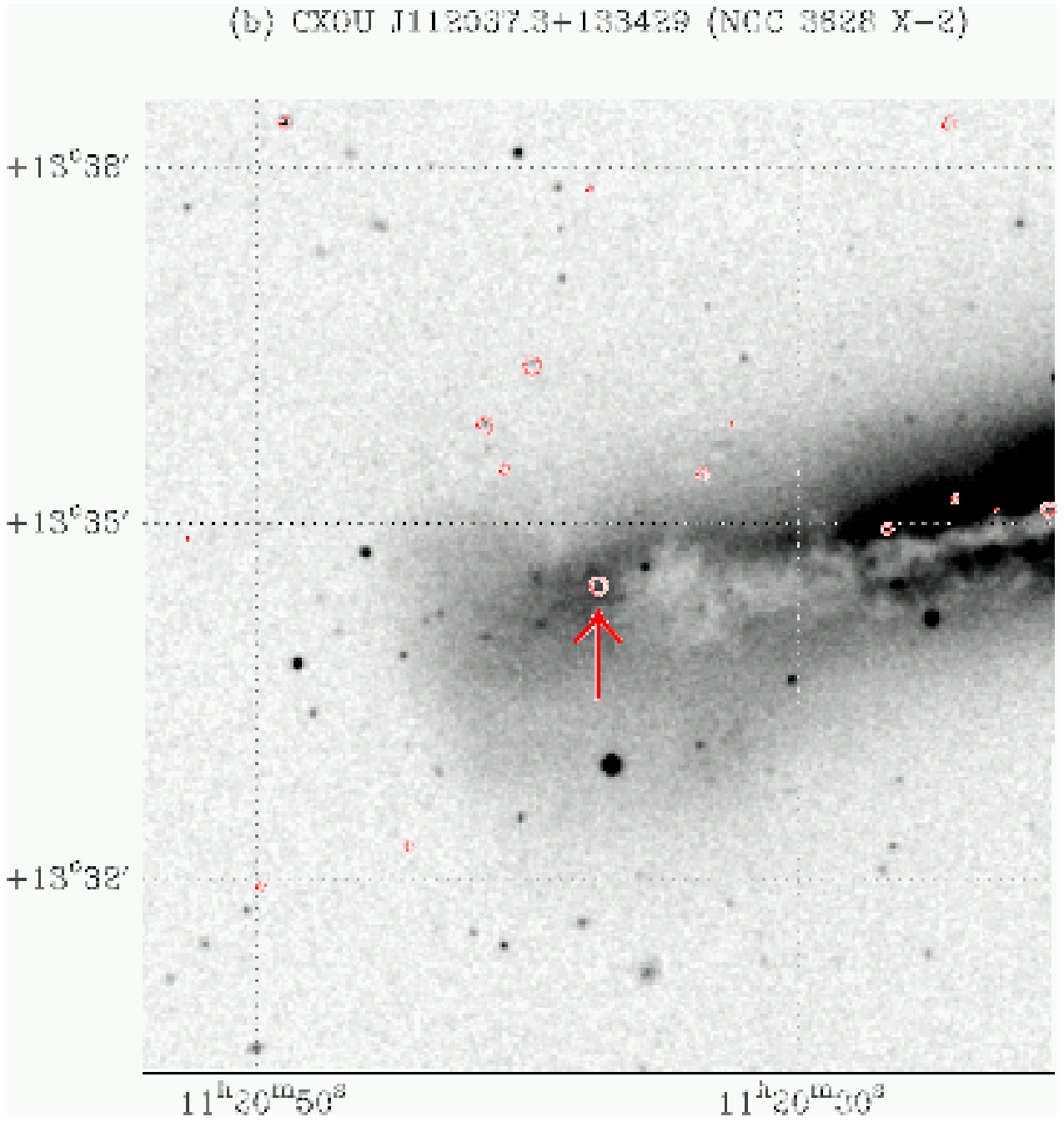}\vspace*{5mm}
\includegraphics[width=8cm]{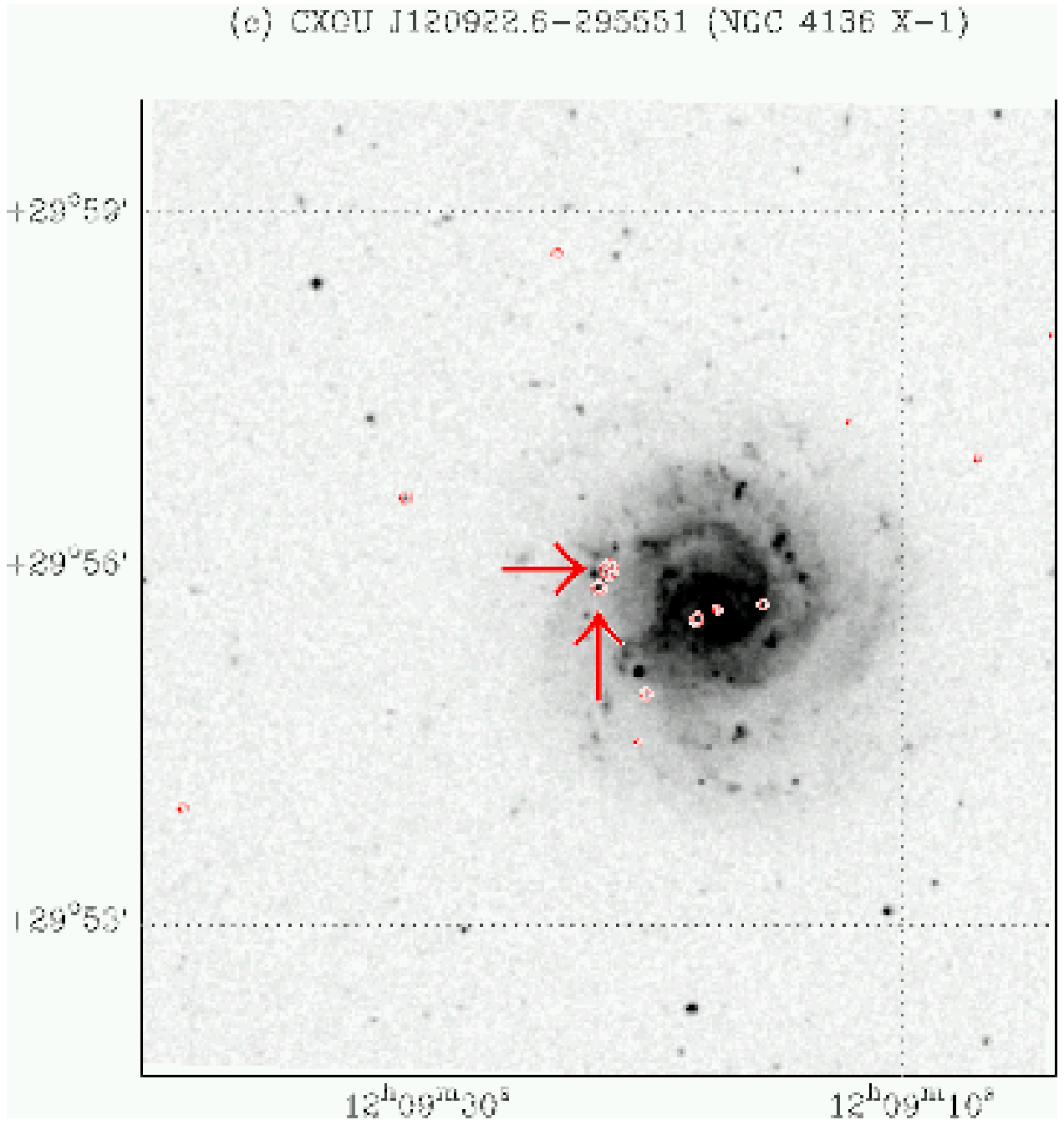}
\hspace*{2mm}
\includegraphics[width=8cm]{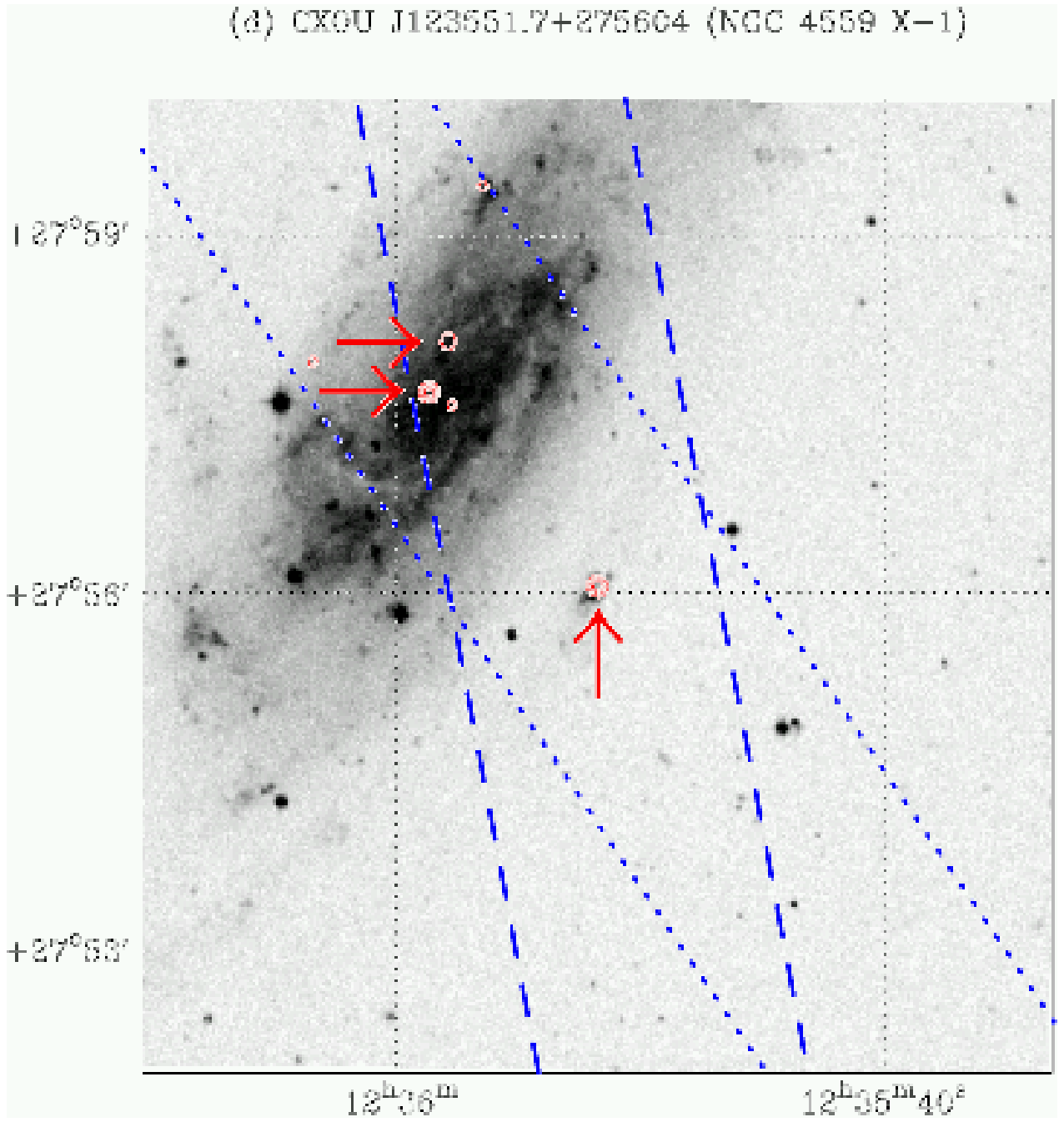}
\caption{The locations of four of the five target ULXs.  Each figure
shows an $8.1' \times 8.1'$ DSS-2 blue image centred on the position
of the ULX, rotated slightly to match the \chan projection.
Coordinates are shown in J2000.  The \chan X-ray emission contours are
overlayed in red to highlight the positions of the X-ray sources in
each field-of-view.  The \chan data has been smoothed by a HWHM 3
pixel Gaussian mask to aid visibility, and the contours are plotted at
0.3 and 10 count pixel$^{-1}$.  The target ULXs are highlighted by the
vertical arrows, and the field ULXs by horizontal arrows.  The spatial
coverage of the \chan sub-arrays used for observations of the IC 342
X-1 and NGC 4559 X-1 fields are highlighted in blue for each
respective field, with the first epoch coverage shown by the dashed
lines, and the second epoch by the dotted lines.}
\label{ulxlocations}
\end{figure*}

\begin{figure*}
\setcounter{figure}{1}
\centering
\includegraphics[width=8cm]{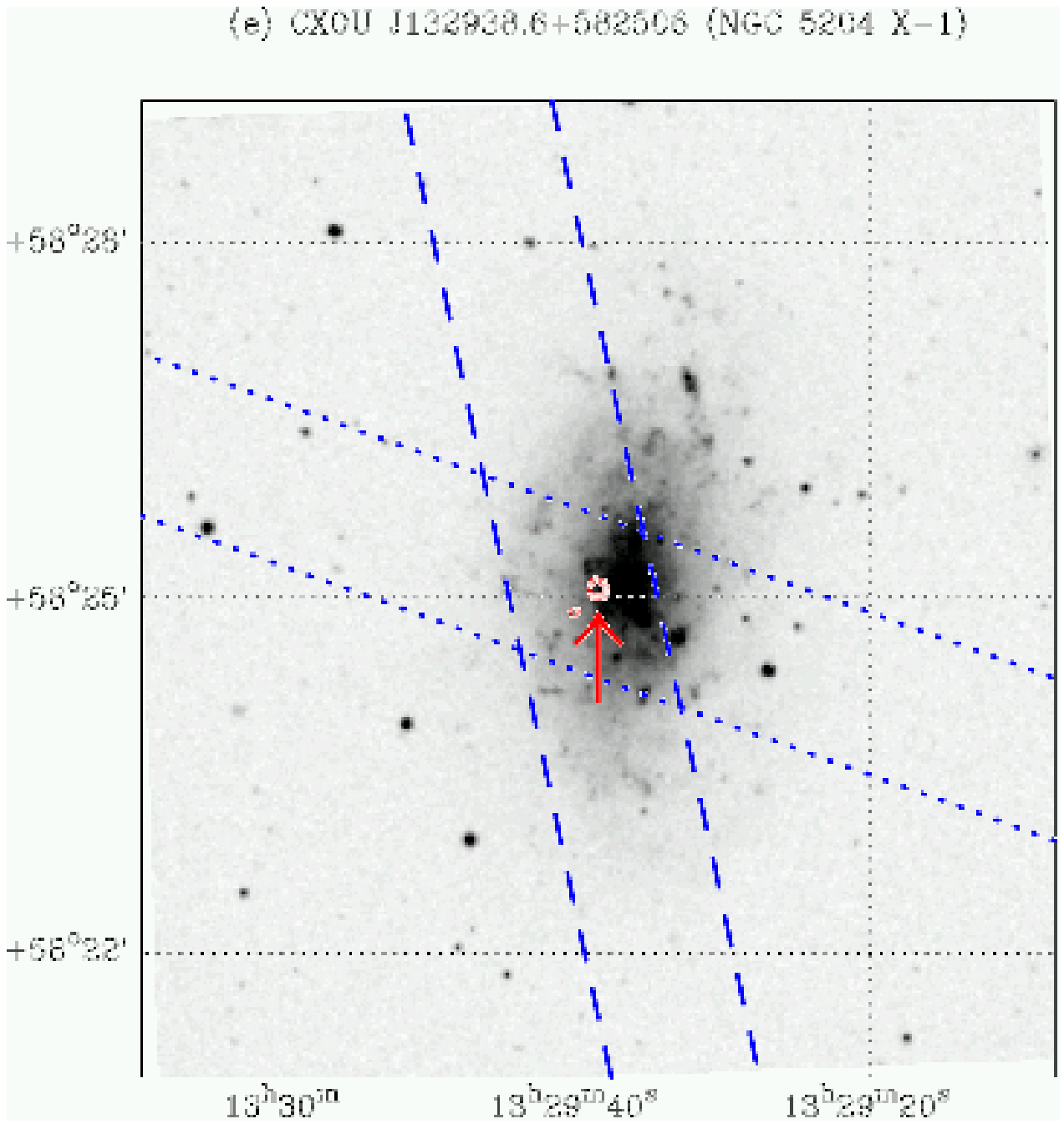}
\caption{As per figure~\ref{ulxlocations}, for the NGC 5204 X-1 field.}
\label{moreulxlocations}
\end{figure*}

If the ULXs in these galaxies are associated with young stellar
populations, we might expect to find them co-located with the regions
of the galaxy most prone to hosting star formation, such as the spiral
arms, whereas if they are associated with the older stars then their
distribution is more likely to be centrally peaked, or spread evenly
throughout the disc.  Table~\ref{galparams} shows that the majority of
the ULXs (five out of eight) are located more than halfway out from
the nucleus towards the edge of the galaxy.  Four out of these five
appear coincident with spiral arms; the fifth (CXOU J112037.3+133429)
is located in an edge-on system where we cannot distinguish whether it
is in an arm, or an inter-arm region.

The identification of four of the ULXs with spiral arms is relatively
robust, as in both IC 342 and NGC 4136 the galaxies are face-on and so
line-of-sight confusion through the galaxy is minimised.  This is also
true for CXOU J123551.7+275604 in NGC 4559, as this is located in a
faint outer spiral arm clearly separated in projection from the main
body of the galaxy.  The co-location of the ULXs with spiral arms
itself argues that they could be associated with a young stellar
population.  If they are ordinary X-ray binaries, with typical kicks
imparted from the formation of the compact primary in a supernova
explosion of $\sim 100 \kmsec$ (c.f. recent results for GRO 1655-40,
Mirabel et al. 2002), then in $10^7$ years this would move the
position of the ULX relative to its birth place by a maximum of 1 kpc,
consistent with an observed position in or near a spiral arm if born
there.  However, if the ULX are associated with an older population
then in their $> 10^8$ year lifetime they could be displaced by 10 kpc
or more, clearly leaving no requirement for an association with the
spiral arms.  The arguments in favour of these ULXs being related to
young stellar populations are supported by follow-up observations of
the immediate environments of three of these ``outer'' ULXs, with \hii
regions detected close to (within 200 pc of) CXOU J123551.7+275604 and
coincident with CXOU J120922.6+295551 (Roberts et al., in prep.), and
the detection of a possible supernova remnant encircling the position
of CXOU J034555.7+680455 (Roberts et al. 2003).  However, the kick
argument above requires that either the ULXs are very young, or
possess a small kick velocity and/or a velocity almost entirely
projected along our line-of-sight for these to be true physical
associations rather than line-of-sight coincidences with the star
formation regions we might expect to observe within a galaxy's spiral
arms.

Line-of-sight confusion is a much bigger issue for the three ULXs that
are located closer to the centre of their respective host galaxies.
Despite this, detailed follow-up of CXOU J132938.6+582506 reveals it
to be associated with young stars (Goad et al. 2002).  The remaining
two sources are located close to the centre of NGC 4559.  However, the
nucleus of this galaxy is known to host active star formation (Ho,
Filippenko \& Sargent 1997), implying that it is possible (though by
no means certain) that these ULXs may also be related to the presence
of young stars.  It is entirely plausible, then, that all the ULXs in
this small sample may be associated with young stellar populations.
This is consistent with previous studies finding comparatively large
populations of ULX in the most active star forming galaxies
(e.g. Fabbiano, Zezas \& Murray 2001; Lira et al. 2002; Zezas, Ward \&
Murray 2003), indicating a relationship between many ULX and active
star formation, though interestingly in this case we co-locate them
with the star forming regions of otherwise relatively normal galaxies.

To put this in context, it is worth noting that the host galaxies in
our sample are all type Sb or later, indicating that they are
disc-dominated, and so do not possess large bulges dominated by old
stars.  We can investigate the ULX -- young stellar population link
further by reference to the ULX survey of Colbert \& Ptak (2002).
They detect more than 30 candidate ULXs coincident with, or in the
haloes of, elliptical galaxies, which must be dominated by an old
stellar population.  Averaging over the fourteen elliptical galaxies
in which ULXs are located with respect to the blue luminosity of each
galaxy, L$_B$, gives 0.8 ULX per $10^{10}$ L$_B$.  This is very much
an {\it upper\/} limit on the number of ULX from an old stellar
population, as Colbert \& Ptak search out to a radius of 2R$_{25}$,
potentially allowing a much greater contamination from background
objects than simply focussing within the standard definition of the
galaxy size, the 25 mag arcsec$^{-2}$ isophotal ellipse.  Also, since
many of their candidate ULXs are possible halo objects, these may
therefore constitute a physically separate population to that
associated with the old stellar population.  Nevertheless, by
comparing their average ULX to L$_B$ ratio with the blue luminosities
of our target galaxies (and correcting for the coverage of each galaxy
in the \chan observations), we can establish a conservative upper
limit on the number of our ULXs associated with the older stellar
populations.  This turns out to be an upper limit of two out of the
eight ULXs we observe.  We therefore confirm that our sample is very
likely to be dominated by ULXs associated with a young stellar
population.

\section{X-ray characteristics}

In this section we investigate the detailed X-ray properties of the
sources.  Unless specifically stated, we perform the analyses on all
five target plus all three field ULXs.  Discussion of these
characteristics in terms of physical models is deferred until later
sections.

\subsection{Spatial extent}

A primary goal of this programme was to use the unique 0.5-arcsecond
spatial resolution of \chan to investigate whether these ULXs are
resolved into complexes of many X-ray emitting sources, or whether
they remain a single, point-like object at the highest available X-ray
resolution.  Hence, the target ULXs were placed at the on-axis
aimpoint of the ACIS-S3 chip in each observation to provide the best
possible spatial data.

To investigate the question of spatial extension, the two observations
of each field were combined to enhance the signal-to-noise ratio of
the images.  The excellent spatial precision of the data (1 pixel
$\equiv 0.492$ arcsecond) was maintained by lining-up the peak in
intensity of each target source.  This method was verified by
examining the peaks in fainter sources within the field, which also
lined-up accurately.  We then derived the radial profile of each ULX
and fit its core with a Gaussian function, following the procedure
described in section 4.2.1 of Roberts et al. (2002)\footnote{Some
adjustments to the size and position of the background annuli were
required to eliminate nearby X-ray sources in several fields.  An
example is the NGC 4136 field, in which the two ULXs are separated by
only $\sim 22$ pixels.}.  The resulting fits gave FWHM of between 1.8
-- 2.2 pixels for all on-axis sources, and $\sim 2.4$ pixels for the
2-arcmin off-axis ULXs in NGC 4559, all consistent with the nominal
\chan point-spread function at those positions.  No evidence was found 
for a faint, extended component surrounding the position of any of the
ULXs.  This demonstrates that the ULXs are point-like at the
0.5-arcsecond spatial resolution of {\it Chandra\/}, corresponding to
maximum physical sizes for the X-ray emitting regions of $\sim 9$ --
23 parsecs in the host galaxies.  The spatial data is also, of course,
consistent with all the objects being single, point-like X-ray
sources.

\subsection{Spectral properties}

Spectra were extracted in an eight arcsecond diameter aperture centred
on the position of each ULX using the {\small PSEXTRACT} script, which
also retrieves the appropriate response matrices and ancillary
response files for each observation.  A source-free local background
region of equivalent size was used in each case, though its impact was
of little significance since it typically contained only $0.1 - 1 \%$
of the counts accumulated from the source.  The ancillary response
files were corrected for the gradual in-orbit degradation in the
quantum efficiency of the ACIS detectors using the {\small CORRARF}
tool\footnote{See {\tt
http://asc.harvard.edu/cal/Acis/Cal\_prods/qeDeg}.}.  The spectra were
grouped to a minimum of 20 counts per bin, and then analysed in
{\small XSPEC v.11.2}, excluding data below 0.5 keV due to the
uncertainty in the calibration at these energies.  In the following
analysis, all quoted errors are the 90\% confidence errors for one
interesting parameter.

A major consideration in the design of these observations was to limit
the distorting effect of detector pile-up on the observed X-ray
spectrum of each ULX.  As stated in Section 2, the \ro HRI count rates
of each ULX were used to determine the optimum sub-array size to lower
the anticipated pile-up fraction below 10\%.  However, this did not
guarantee that each source would have an acceptable level of pile-up,
due to the long-term variable nature of the ULXs themselves (see
below). We can derive a lower limit on the observed pile-up fraction
of each observation by reference to Figure 6.24 of the \chan
Proposer's Observatory Guide (v.4)\footnote{See {\tt
http://asc.harvard.edu/proposer/POG/index.html}.}, which shows pile-up
fraction as a function of counts per readout frame (we calculate a
lower limit since our input is the {\it observed\/} counts per frame,
which already includes piled-up events).  This is very much a first
order approximation, since it does not include, for example, the
effect of variations in the source spectra on the degree of pile-up.
Only one observation, the 2001 January observation of CXOU
J132938.6+582506, turns out to have a lower limit in excess of 10\%,
and this is only marginal at 11\%.  All other observations have limits
of 8\% pile-up or less, implying that the policy of using sub-arrays
was successful.

We investigated the presence of residual pile-up effects using the
{\small XSPEC v.11.2} parameterisation of the CCD event pile-up model
of Davis (2001).  A simple absorbed powerlaw continuum model, both
with and without the pile-up model, was fit to the observed spectrum
of the brightest ULXs (namely CXOU J132938.6+582506, CXOU
J123551.7+275604 and CXOU J034555.7+680455 in both epochs, and CXOU
J123558.6+275742 in its second observational epoch).  In the pile-up
model we allowed $\alpha$, the grade morphing parameter, to vary
freely, and set the frame time to 0.7 seconds and 1.1 seconds for the
${1}\over{8}$ and ${1}\over{4}$ sub-arrays respectively, with the
other parameters fixed as default.  In six of the seven cases the
changes to the spectral fit when the pile-up algorithm was applied
were minimal ($\Delta \chi^2 < 3$ for one extra degree of freedom).
Though in most cases the powerlaw photon indices became slightly
softer, as expected since CCD pile-up acts to harden the observed
source spectra, this change was offset by much worse constraints on
the parameters.  Hence, the absorption columns and powerlaw photon
indices were consistent both with and without the inclusion of the
pile-up model, within the derived errors, in all six cases.  None of
these six fits were able to place strong constraints on $\alpha$.  We
therefore consider pile-up effects to be negligible in the X-ray
spectra of all but one of the ULXs.

The only case in which the pile-up model gave a much improved $\chi^2$
was the 2001 January observation of CXOU J132938.6+582506, with
$\Delta \chi^2 = 16.3$ for one extra degree of freedom, and a
statistically significant softening of the spectral slope.  We discuss
this case further below.

\begin{table*}
\caption{Best fits of two simple models to the two-epoch ULX X-ray
spectra.}
\begin{tabular}{llllcllcc}\hline
ULX (CXOU J)	& Epoch &\multicolumn{3}{c}{WA*PO$^a$} &
\multicolumn{3}{c}{WA*DISKBB$^a$}	& \lx$^b$ \\
 & & \nh$^c$	& $\Gamma$$^d$	& $\chi^2$/dof	&
\nh$^c$	& $kT_{in}$$^e$	& $\chi^2$/dof	& \\\hline
034555.7+680455	& 2002-04-29	& $0.52 \pm 0.07$	&
$1.63^{+0.13}_{-0.12}$	& {\bf 78.8/81} & $0.30^{+0.05}_{-0.04}$ &
$1.81^{+0.22}_{-0.18}$	& 97.4/81 & 4.4(5.9) \\
	& 2002-08-26	& $0.61 \pm 0.08$	&
$1.70^{+0.12}_{-0.13}$	& {\bf 92.0/87} & $0.36 \pm 0.05$ &
$1.76^{+0.19}_{-0.16}$	& 105.7/87 & 4.4(6.4) \\
112037.3+133429	& 2002-04-06	& 0.022$^f$	&
$1.57 \pm 0.24$	& {\bf 8.4/10} & $0.022^f$ &
$0.95^{+0.3}_{-0.2}$	& 12.3/10 & 0.7(0.8) \\
	& 2002-07-04	& $0.16^{+0.09}_{-0.13}$	&
$2.20^{+0.34}_{-0.19}$	& {\bf 4.4/10} & $0.022^f$ &
$0.86^{+0.18}_{-0.14}$	& 6.4/11 & 0.6(0.7) \\
120922.2+295600	& 2002-03-07	& 0.13$^{+0.08}_{-0.07}$	&
$1.68^{+0.27}_{-0.22}$	& {\bf 30.1/19} & $0.016^f$ &
$1.46^{+0.27}_{-0.25}$	& 37.7/20 & 2.3(2.6) \\
	& 2002-06-08	& $< 0.21$	&
$1.55^{+0.30}_{-0.33}$	& {\bf 13.9/13} & $0.016^f$ &
$1.43^{+0.44}_{-0.24}$	& 18.4/14 & 1.7(1.9) \\
123551.7+275604	& 2001-01-14	& 0.015$^f$	&
$1.91 \pm 0.09$	& {\bf 69.9/52} & $0.015^f$ &
$[\sim 0.84]^g$	& 169/52 & 9.8(10.0) \\
	& 2001-06-04	& $0.04 \pm 0.03$	&
$2.16^{+0.15}_{-0.14}$	& {\bf 100/72} & $0.015^f$ &
$[\sim 0.66]^g$	& 222/73 & 11.6(12.5) \\
123558.6+275742	& 2001-01-14	& $0.17^{+0.08}_{-0.07}$	&
$1.98^{+0.23}_{-0.24}$	& 26.3/22 & $0.015^f$ &
$1.14^{+0.15}_{-0.13}$	& {\bf 24.6/23} & 4.2(4.3) \\
	& 2001-06-04	& $0.16 \pm 0.06$	&
$1.82^{+0.17}_{-0.15}$	& 63.6/51 & $0.015^f$ &
$1.30^{+0.12}_{-0.10}$	& {\bf 49.2/52} & 8.9(9.1) \\
132938.6+582506	& 2001-01-09	& $0.10 \pm 0.02$	&
$2.38^{+0.10}_{-0.11}$	& {\bf 141/109} & $0.014^f$ &
$[\sim 0.7]^g$	& 260/110 & 5.7(6.9) \\
	& 2001-05-02	& $0.10^{+0.05}_{-0.04}$	&
$2.96^{+0.25}_{-0.21}$	& {\bf 41.1/46} & $0.014^f$ &
$0.44^{+0.04}_{-0.03}$	& 86.5/47 & 1.8(2.4) \\
\hline
\end{tabular}
\begin{tabular}{l}
Notes: $^a$Spectral model components are shown as per the XSPEC
syntax, with ``WA'' representing a cold absorption model, ``PO'' a
\\powerlaw continuum and ``DISKBB'' the MCDBB model.  $^b$Observed
luminosity in the 0.5 -- 8 keV band in units of $10^{39} \ergsec$.
\\Figures in parentheses give the intrinsic (unabsorbed) luminosity.  We
calculate these values using the best-fitting model highlighted \\by
showing its $\chi^2$/dof in bold.  $^c$Absorption column, in units of
$10^{22} \atpcm$.  $^d$Powerlaw photon index.  $^e$Inner accretion
disc \\temperature in keV.  $^f$Value fixed at the foreground Galactic 
absorption column (see Table~\ref{galparams}).  $^g$Parameter value
not constrained by\\ model fit.
\end{tabular}
\label{xspecfits1}
\end{table*}

\begin{table*}
\caption{Two-component fits to the ULX X-ray spectra.}
\begin{tabular}{lllllllcc}\hline
ULX (CXOU J)	& Epoch	& Model$^a$	& $\alpha$	& \nh	&
$kT/kT_{in}$	& $\Gamma$	& $\chi^2/dof$	& $\Delta \chi^2$$^b$
\\\hline 
120922.2+295600	& 2002-03-07	& WA*(DISKBB+PO)	& - &
$0.81^{+0.37}_{-0.34}$	& $0.12^{+0.05}_{-0.04}$	&
$1.77^{+0.36}_{-0.34}$ 	& 16.7/17 & 13.4 \\
123551.7+275604	& 2001-01-14	& WA*(DISKBB+PO)	& - &
$< 0.25$	& $0.20^{+0.15}_{-0.07}$	&
$1.71^{+0.23}_{-0.16}$ 	& 65.0/49 	& 4.9 \\
	& 2001-06-04	& WA*(DISKBB+PO)	& - &
$0.19^{+0.08}_{-0.14}$	& $0.15^{+0.10}_{-0.03}$	&
$2.10^{+0.18}_{-0.26}$ 	& 92.4/70 	& 7.6 \\
	&	& WA*(MEKAL+PO)	& - & $0.41^{+0.13}_{-0.19}$	&
$0.18^{+0.05}_{-0.02}$	& $2.34^{+0.10}_{-0.14}$ 	& 76.7/70
& 23.3 \\
132938.6+582506	& 2001-01-09	& PILEUP*WA*PO	&
$0.57^{+0.32}_{-0.18}$ & $0.14 \pm 0.03$	& - &
$2.79^{+0.16}_{-0.14}$ & 124/108	& 16.3 \\
\hline
\end{tabular}
\begin{tabular}{l}
Note: $^a$The model components and parameters are as per
Table~\ref{xspecfits1}, except for the MEKAL component which is a
thermal plasma\\ model with its metallicity fixed at solar abundance,
and PILEUP which is as described in the text.  $^b$Improvement in
the\\ $\chi^2$ statistic over the single component model best-fit, for
two extra degrees of freedom (one extra in the pile-up model).
\end{tabular}
\label{complexfits}
\end{table*}

The ULX observations in which less than 250 source counts were
accumulated were not considered for spectral analysis, which ruled out
both observations of CXOU J120922.6+295551 and CXOU J123557.8+275807.
The remaining spectra were initially fit with simple absorbed single
component spectral models, namely a powerlaw continuum, a thermal
bremsstrahlung model, the multi-colour disc blackbody model (hereafter
MCDBB) used to describe an accretion disc around a black hole in its
high (soft) state (Mitsuda et al. 1984), a {\small MEKAL} optically
thin thermal plasma model, and a conventional blackbody spectrum.  The
best-fit parameters to the powerlaw continuum and MCDBB models are
listed in Table~\ref{xspecfits1}, and individual cases are discussed
below.  The thermal bremsstrahlung model provided an adequate fit in
most cases, though never as statistically acceptable as a powerlaw
continuum (or MCDBB in the case of CXOU J123558.6+275742).  We omit it
from the table in favour of the two models that have more generally
been used to describe ULX spectra in past analyses.  The spectra were
generally too hard to provide meaningful fits to the {\small MEKAL}
model, whose parameters tended towards those of the thermal
bremsstrahlung fit.  Similarly, simple blackbody models did not
provide a good fit to any of the spectra.

\begin{figure*}
\centering
\includegraphics[width=6.5cm]{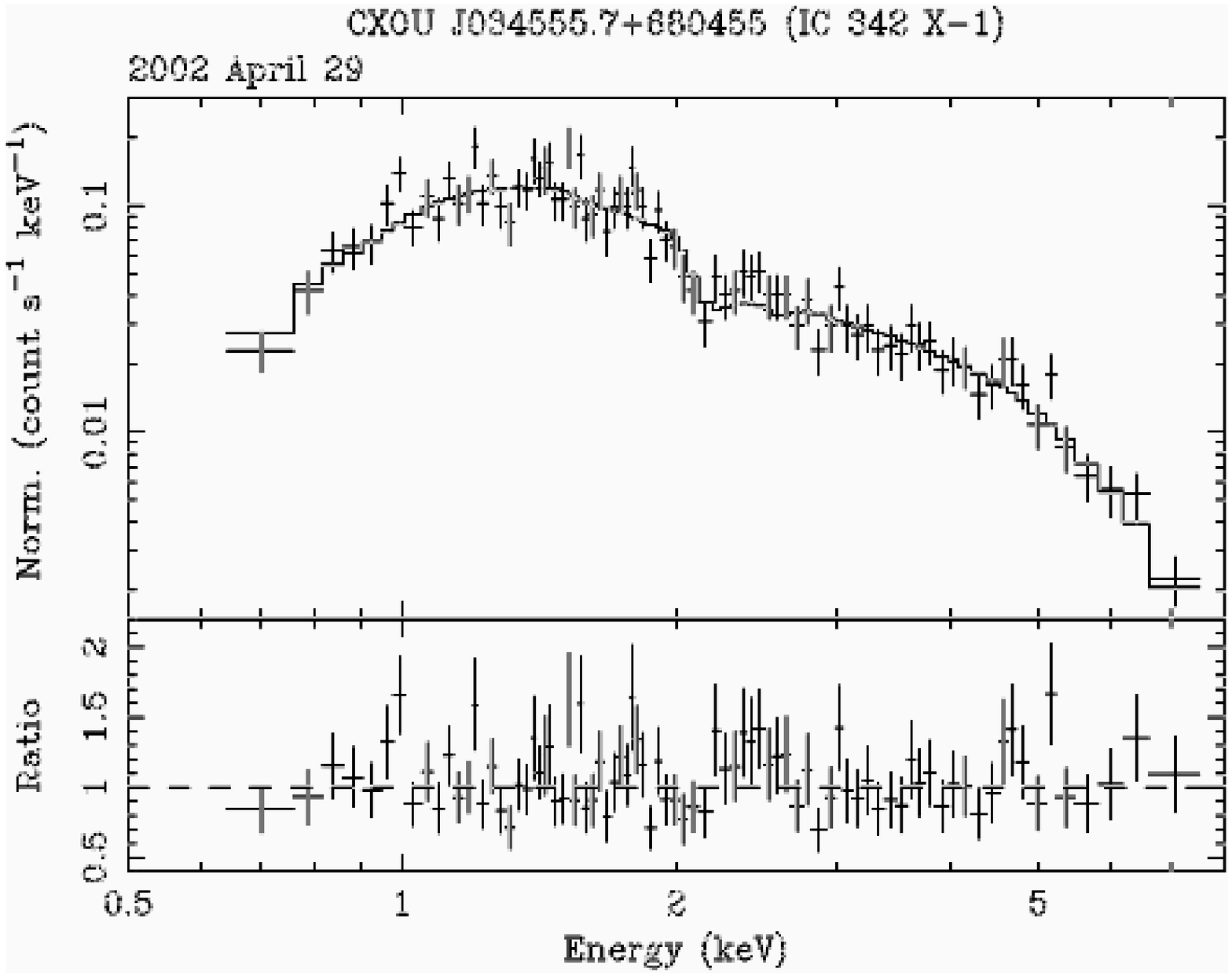}
\hspace*{1.5cm}		
\includegraphics[width=6.5cm]{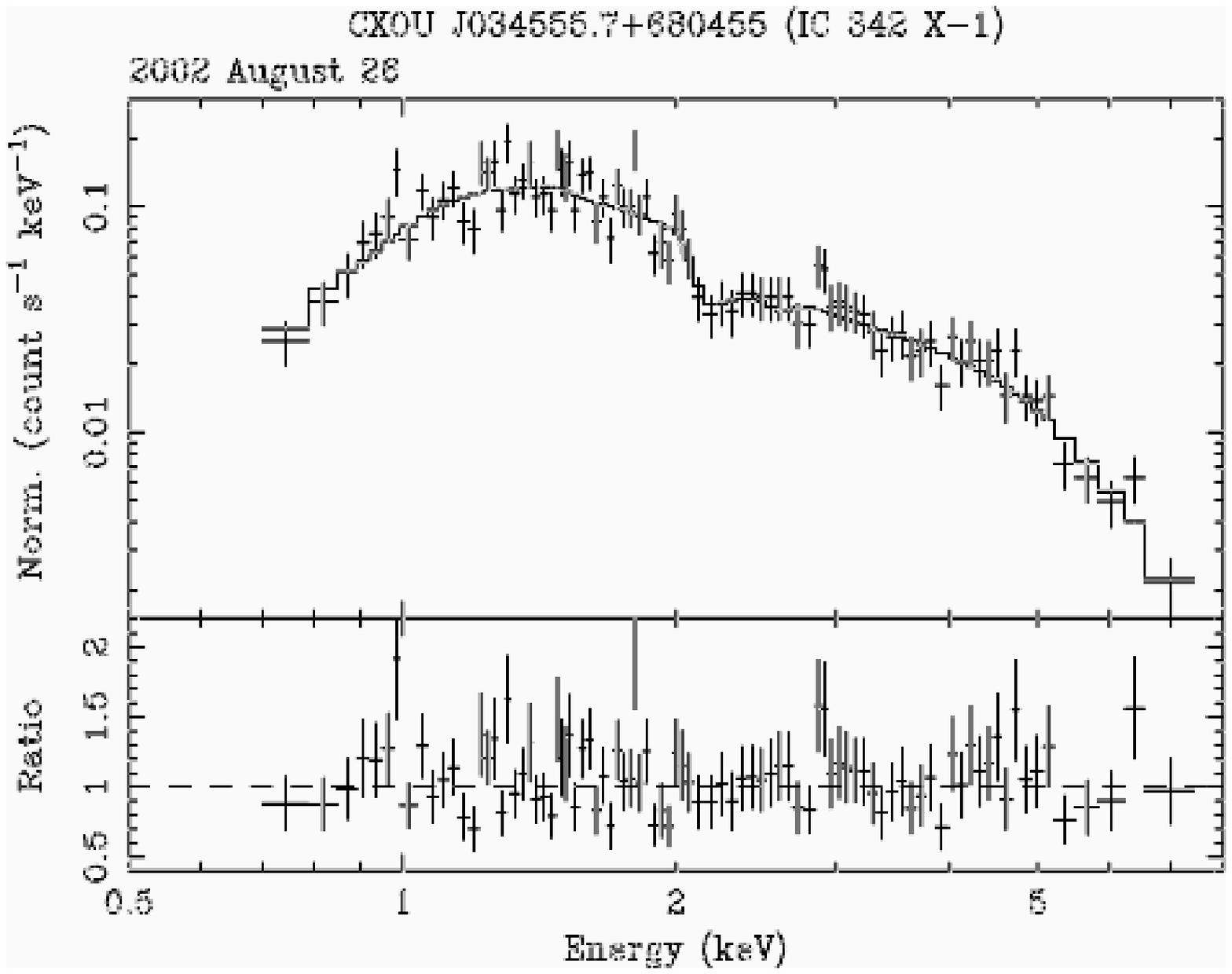}\vspace*{0.3cm}
\includegraphics[width=6.5cm]{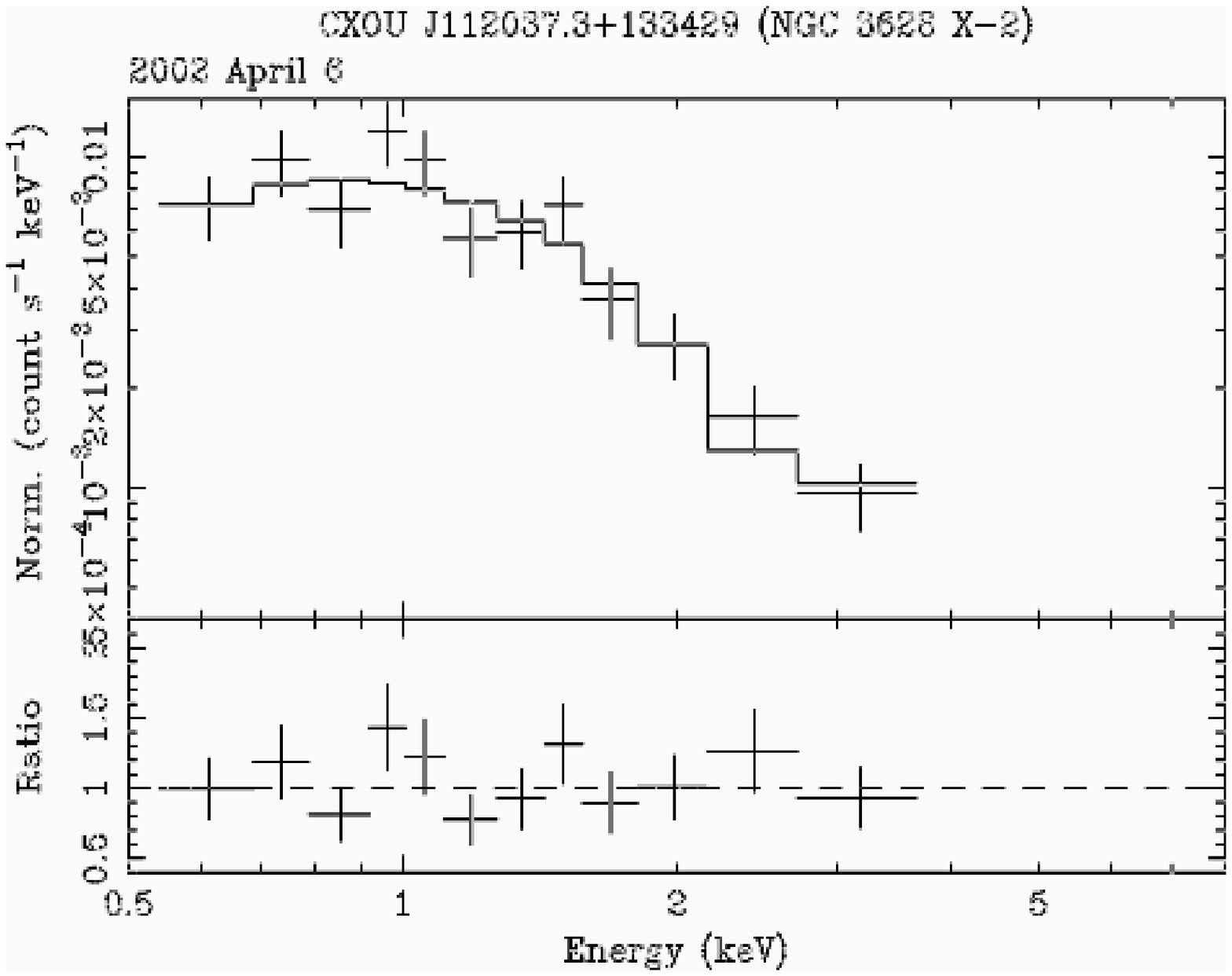}
\hspace*{1.5cm}		  
\includegraphics[width=6.5cm]{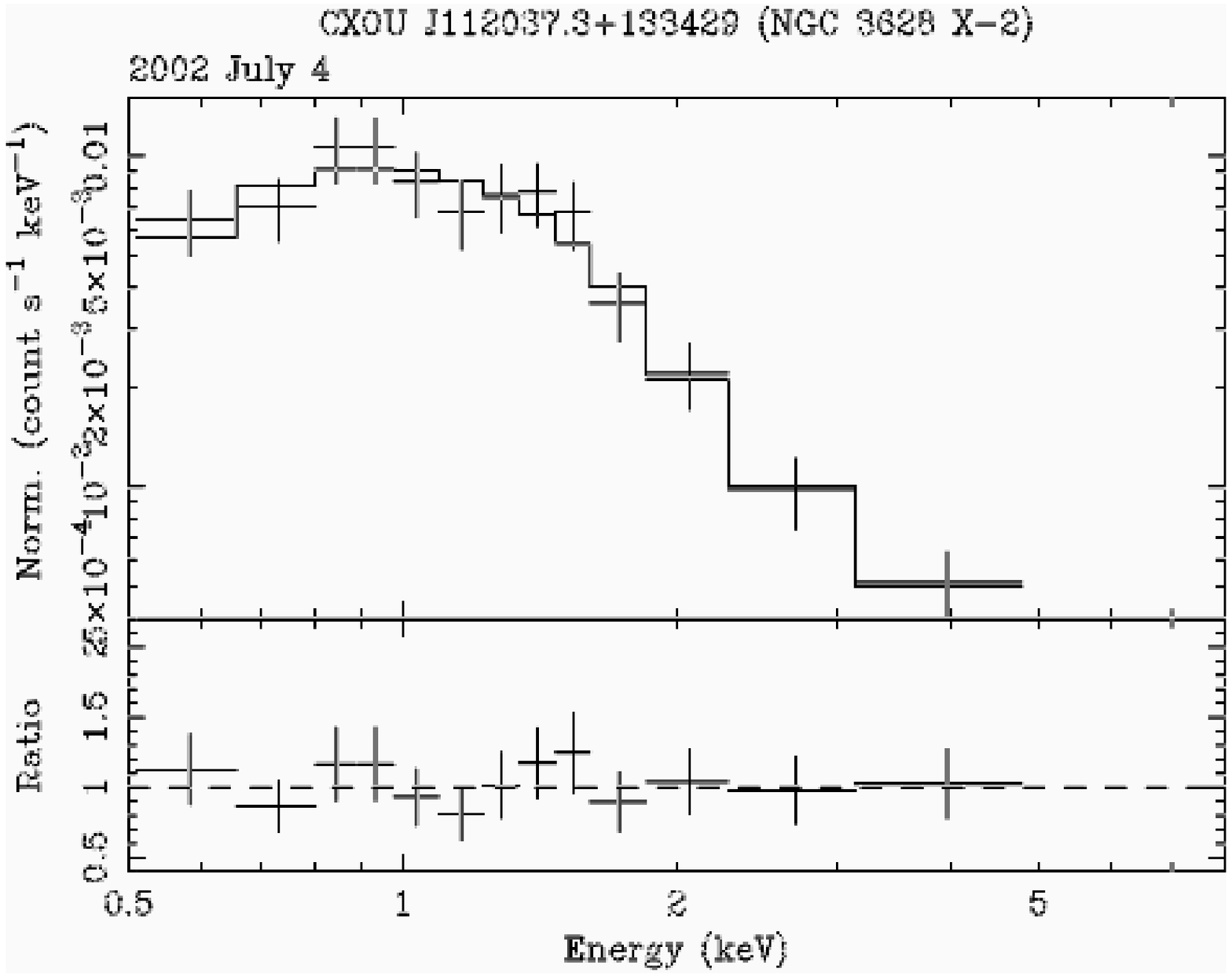}\vspace*{0.3cm}
\includegraphics[width=6.5cm]{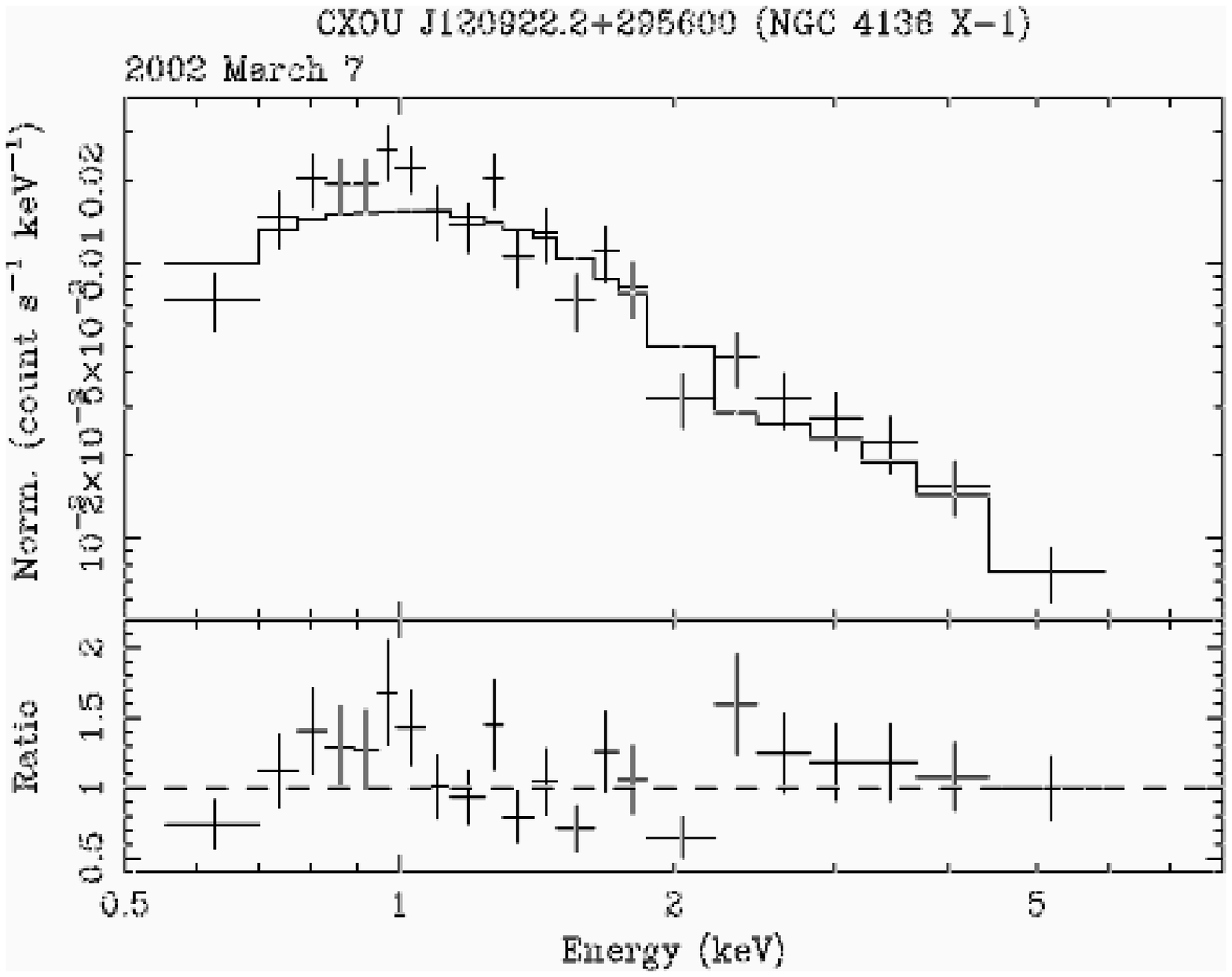}
\hspace*{1.5cm}		  
\includegraphics[width=6.5cm]{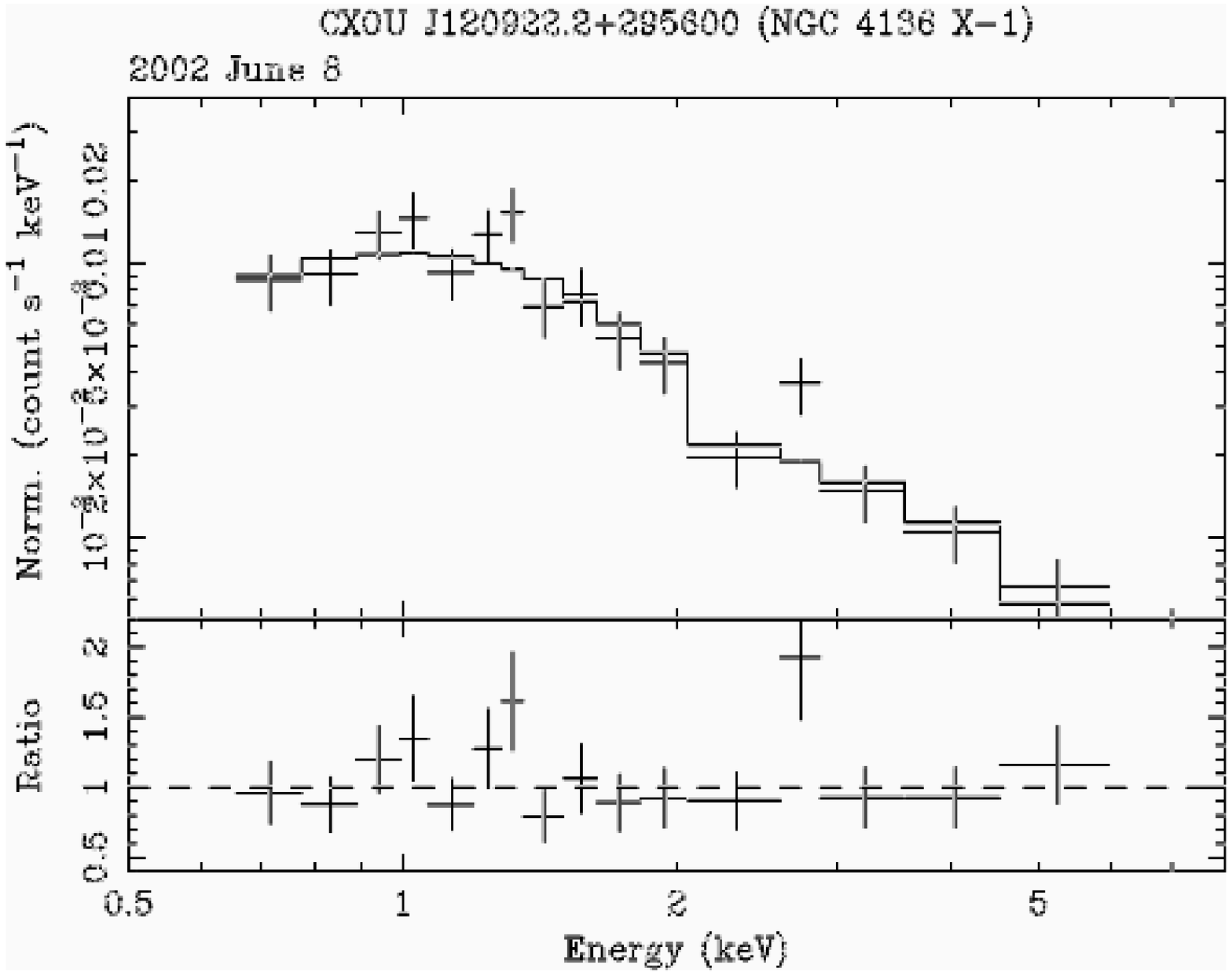}
\caption{The \chan ACIS-S3 spectra of the ULXs.  We show the data and
best-fit single component model in the top window of each panel, and
the ratio of the data to the model in the lower window.  We plot the
first and second epoch data for each ULX in the left and right columns
respectively, using the same normalisation scaling for each pair of
spectra to aid their direct comparison.  All spectra are shown using
the same energy scale, and the ratio values are all displayed over the
same range.}
\label{specs}
\end{figure*}

\begin{figure*}
\centering
\includegraphics[width=6.5cm]{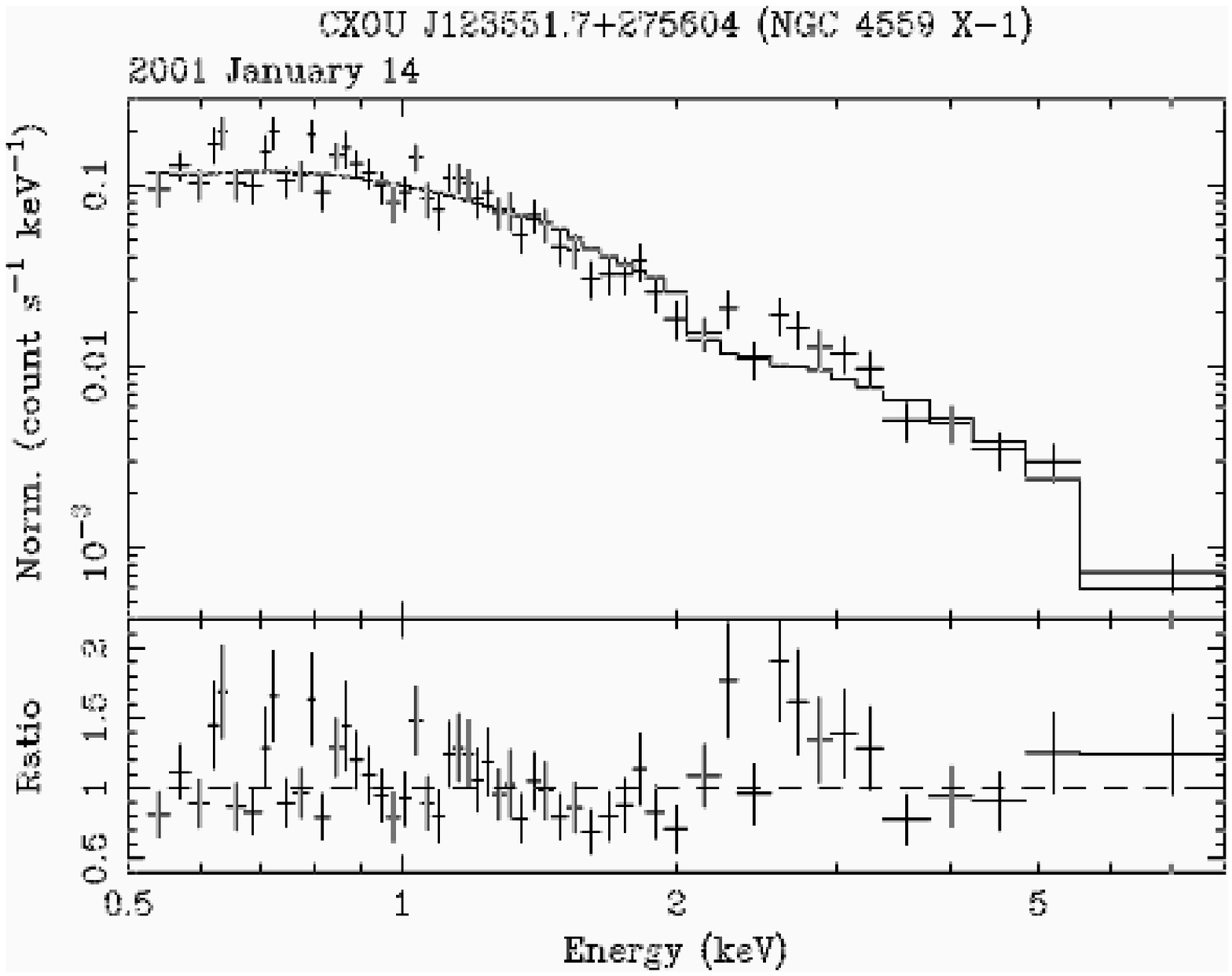}
\hspace*{1.5cm}		    
\includegraphics[width=6.5cm]{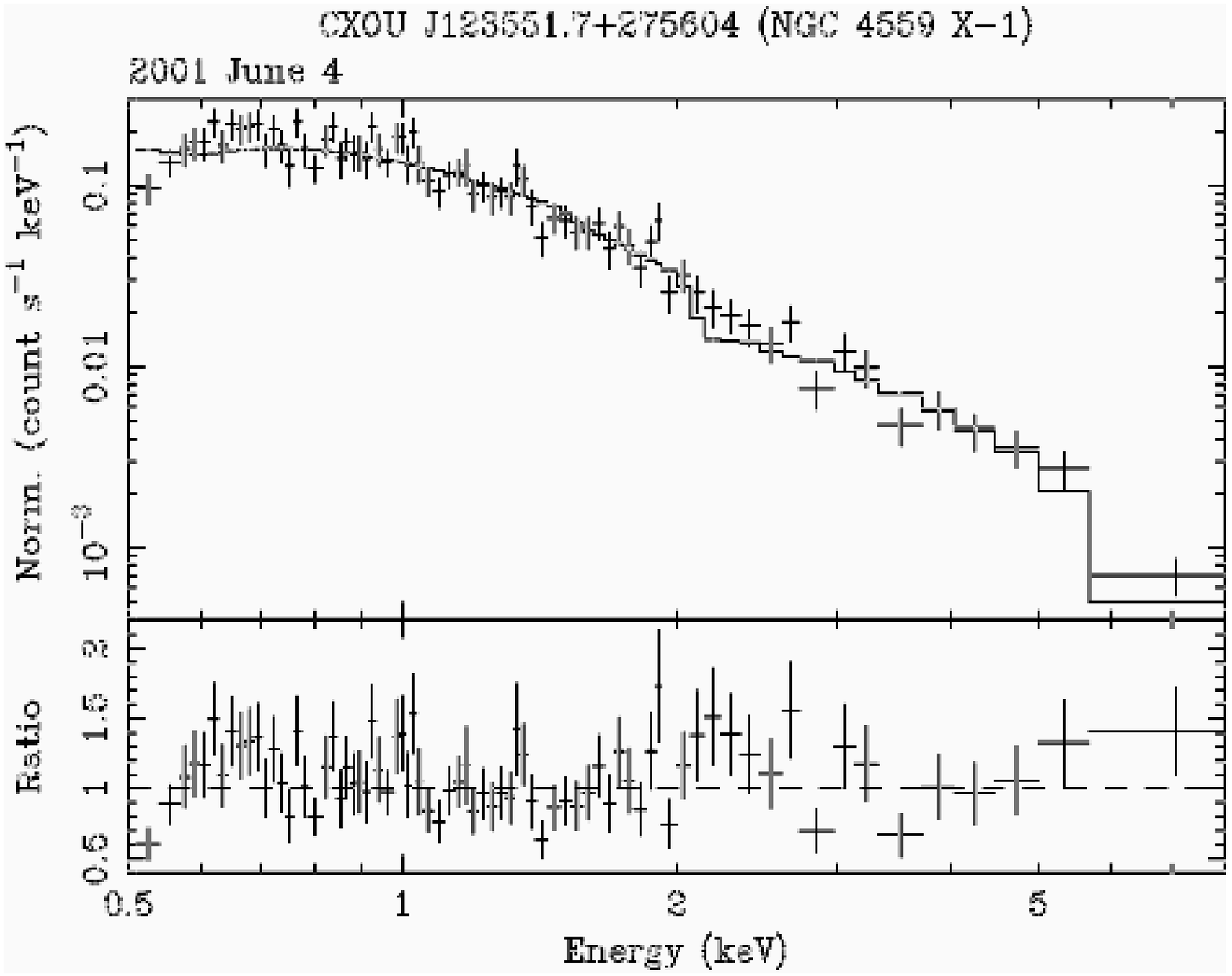}\vspace*{0.3cm}
\includegraphics[width=6.5cm]{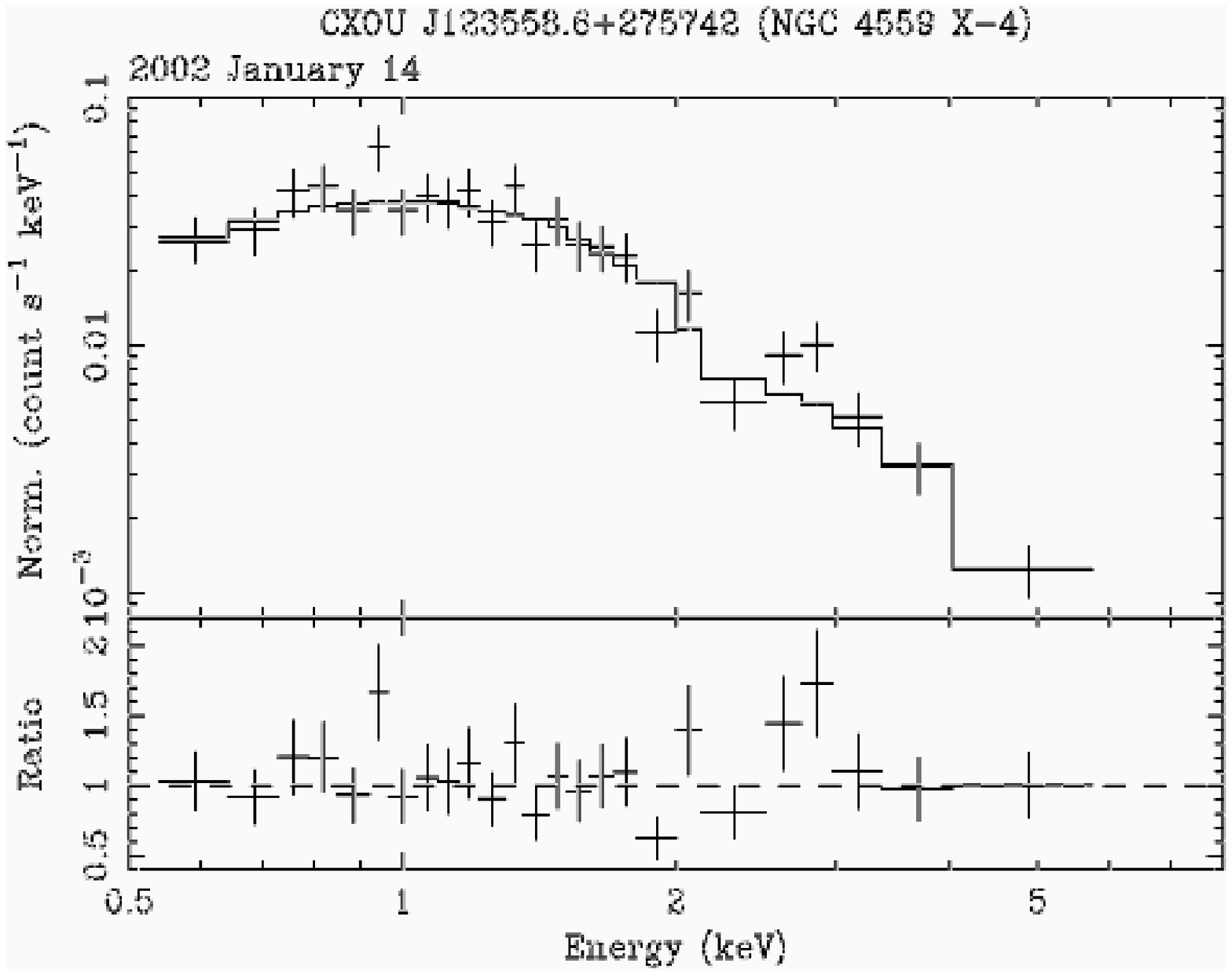}
\hspace*{1.5cm}		    
\includegraphics[width=6.5cm]{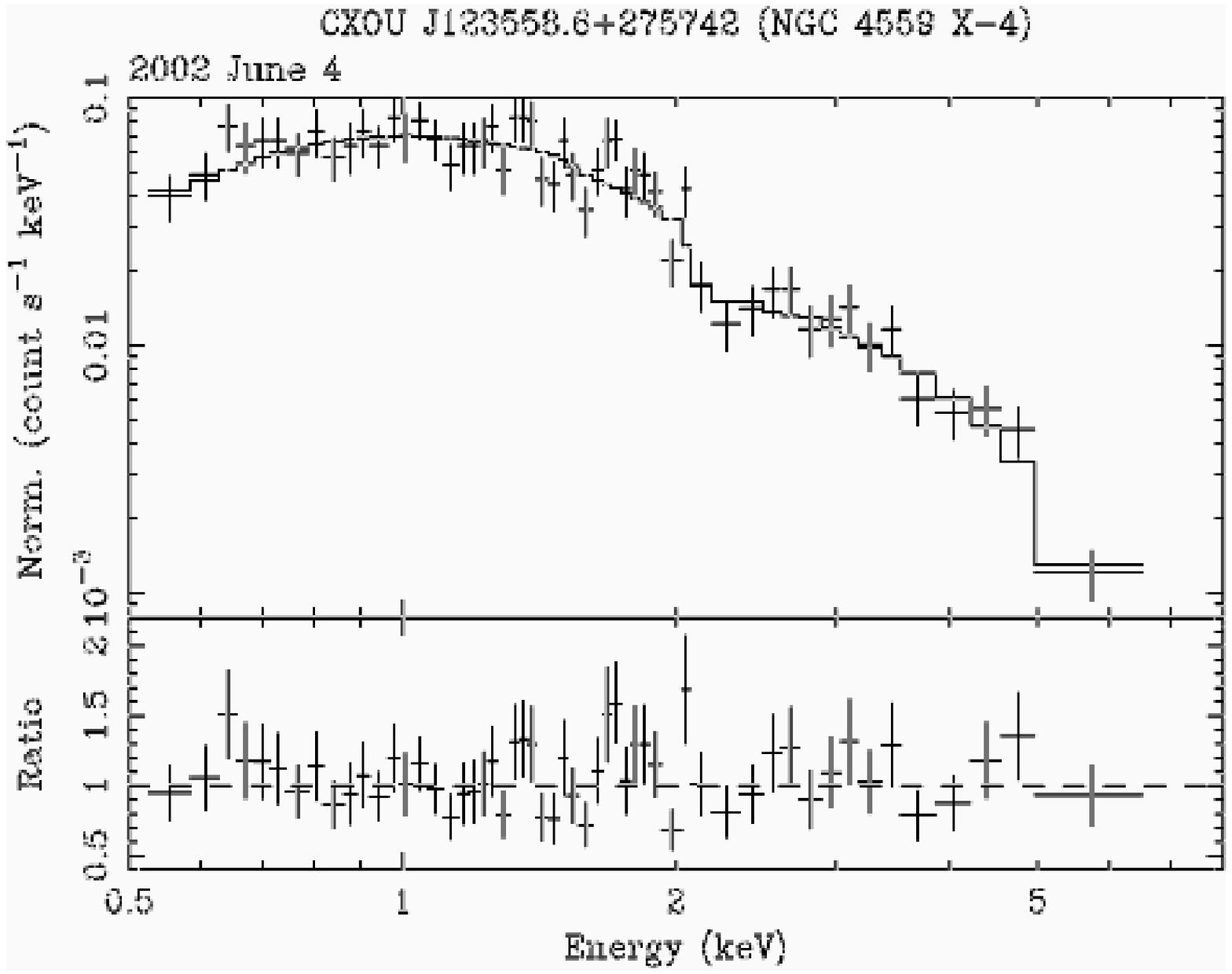}\vspace*{0.3cm}
\includegraphics[width=6.5cm]{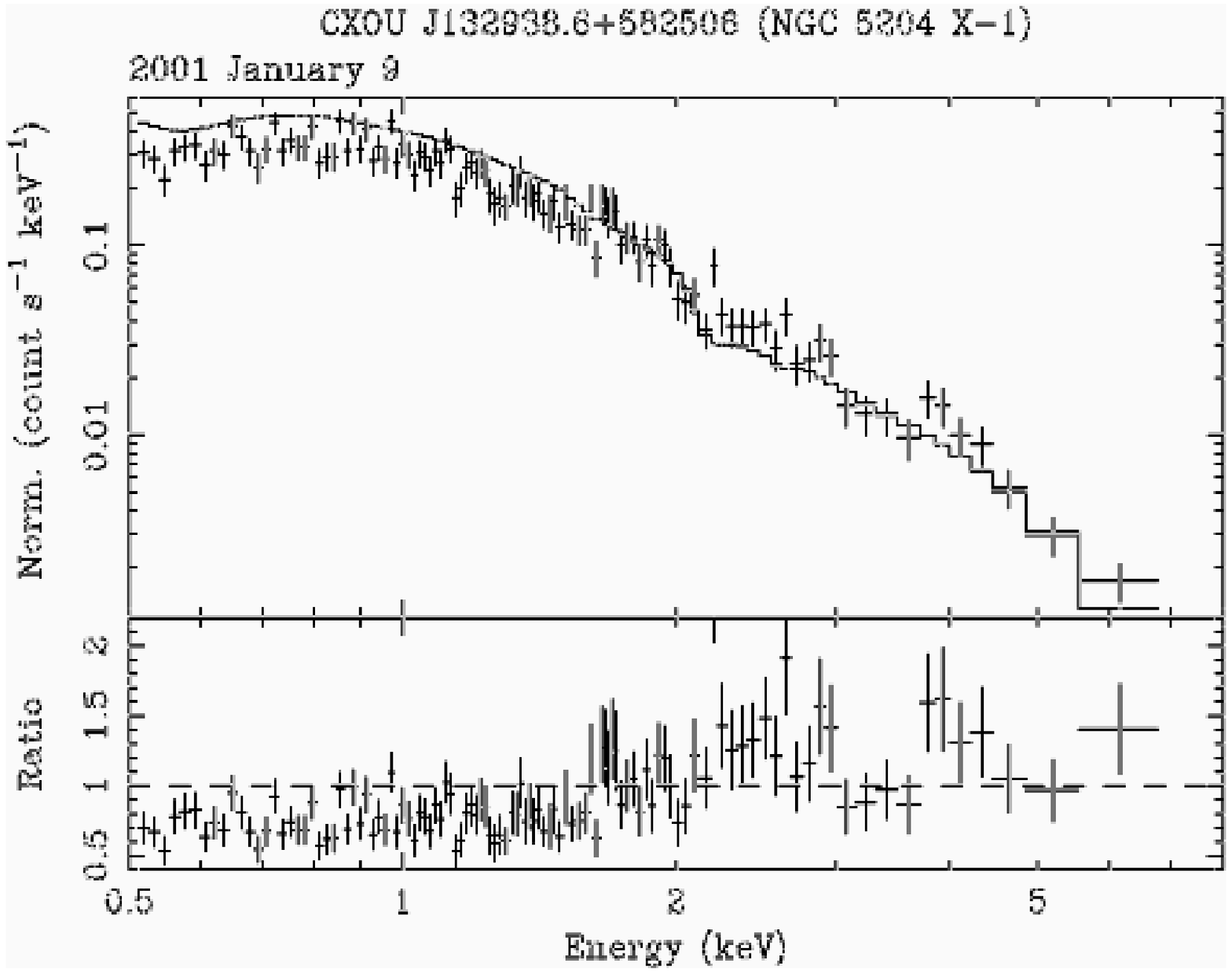}
\hspace*{1.5cm}		    
\includegraphics[width=6.5cm]{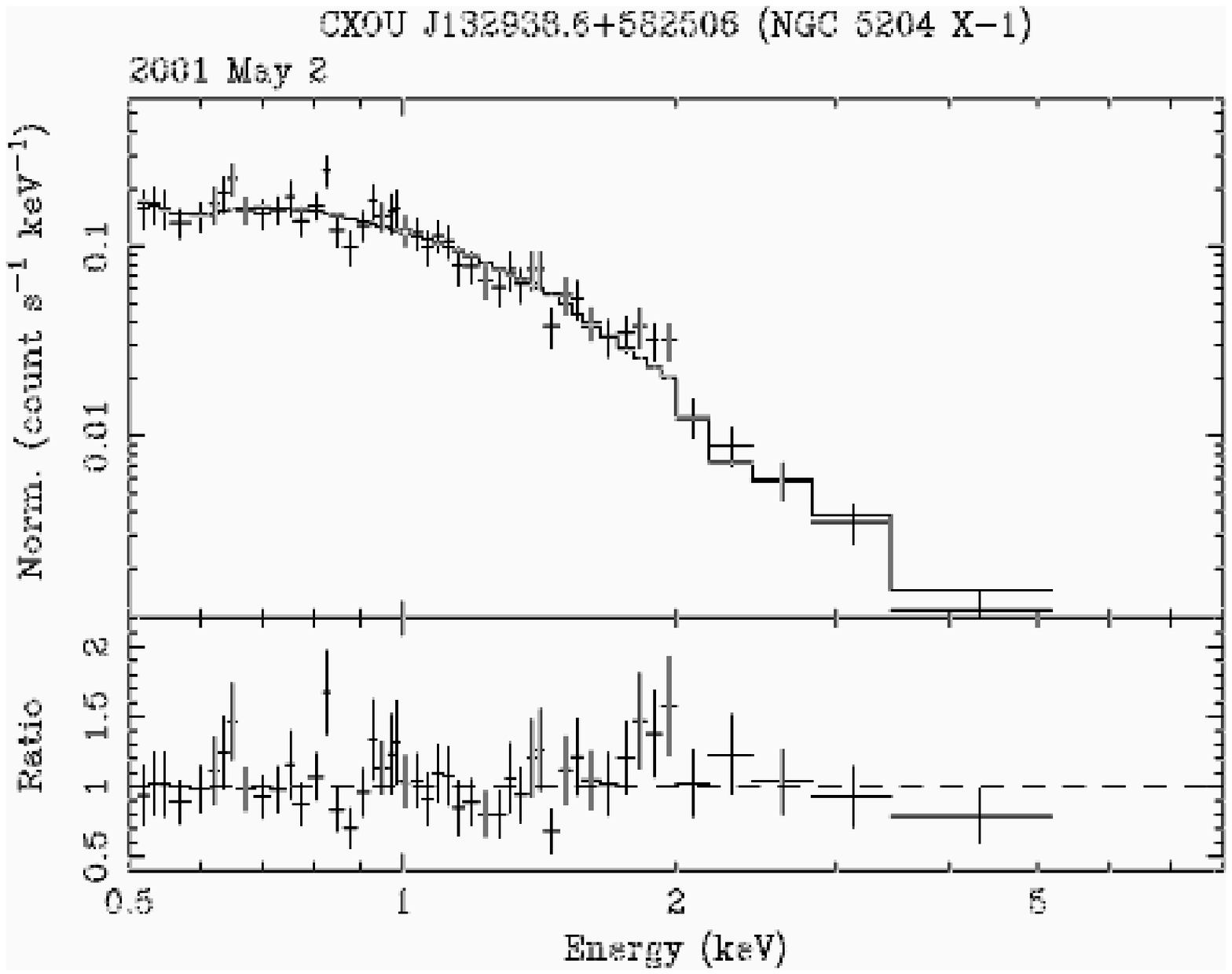}
\caption{As for Figure~\ref{specs}, except for the first epoch
spectrum of CXOU J132938.6+582506.  Here we plot the pile-up
corrected spectral model over the actual data, to demonstrate the
degree of pile-up in the observed X-ray spectrum.}
\label{specs2}
\end{figure*}

As shown in Table~\ref{xspecfits1}, most datasets are adequately
described by a single (absorbed) spectral model component, with a
reduced-$\chi^2$ value at, or below, unity.  However, in several
instances the reduced-$\chi^2$ is still considerably above one.  We
have attempted to fit these datasets with more complex two-component
models (see Table~\ref{complexfits}).  All three datasets for which
the spectral fit was improved by the addition of a second spectral
component (excluding the pile-up correction to CXOU J132938.6+582506)
required the presence of a very soft spectral component.  These are
discussed on a case-by-case basis in the following sub-sections, where
we discuss the spectral fits to each individual ULX.  We demonstrate
the spectral quality of each observation of each source in
Figures~\ref{specs} and~\ref{specs2}.  These show the data and the
best-fit single component model in the upper window of each panel, and
the residuals when the data is divided by the best-fitting model in
the lower panel.

\subsubsection{CXOU J034555.7+680455 (IC 342 X-1)}

This ULX showed a fairly constant X-ray spectrum between the two
observations, with an absorbed powerlaw continuum providing a good fit
to the data from both epochs.  In each case the absorption was roughly
twice the foreground value (c.f. Table~\ref{galparams}), with the
additional absorption of 2 -- 3 $\times 10^{21} \atpcm$ in excess of
the integrated column through IC 342 at the ULX position ($\sim 8
\times 10^{20} \atpcm$; Crosthwaite, Turner \& Ho 2000).  This implies 
a source of additional absorption intrinsic to, or in the environment
of, CXOU J034555.7+680455.  This may originate in the nebula
surrounding this ULX described by Roberts et al. (2003).  The
\chan spectra are consistent with the low/hard spectral state
observed by \asca in 2000 February 24 - March 1, described by an
absorbed powerlaw continuum with \nh $= 0.64 \pm 0.07 \times 10^{22}
\atpcm$ and $\Gamma = 1.73 \pm 0.06$ (Kubota et al. 2001).  This state
has also been interpreted as an anomalous/very high state (Kubota,
Done \& Makishima 2002), which we discuss further in Section 5.1.  The
luminosity also appears little changed, at $\sim 5 \times 10^{39}
\ergsec$ (versus $6 \times 10^{39} \ergsec$ in \asca) when
extrapolated to the 0.5 -- 10 keV band.  Hence it appears that this
ULX has been in a constant spectral state in three separate
observations over 2.5 years.

\subsubsection{CXOU J112037.3+133429 (NGC 3628 X-2)}

This ULX had faded to a luminosity of only $\sim 7 \times 10^{38}
\ergsec$ in our \chan observations, hence the spectra are of low
quality, and the spectral models were not strongly constrained.  A
powerlaw continuum was again the preferred model in both epochs,
though the MCDBB also gave a statistically-acceptable fit in the
second epoch.  The source flux appeared to change little between the
epochs, perhaps fading slightly in the three months between
observations.  However, despite the low quality, the intrinsic X-ray
spectrum showed changes between the epochs, with the first observation
showing an intrinsically hard ($\Gamma \sim 1.57$) but unobscured
powerlaw continuum, whereas the second observation showed a softer
intrinsic slope ($\Gamma \sim 2.2$) with a low-energy turnover due to
absorption.

\subsubsection{CXOU J120922.2+295600}

This new ULX has a spectrum which is best-fit by a powerlaw continuum
model in both epochs.  However, the powerlaw fit is only
statistically-acceptable in the second epoch.  In order to improve the
fit to the first epoch data we tried a variety of two-component
spectral fits to the data.  The best fits came from models with a
highly absorbed, very soft component in addition to a hard powerlaw
continuum.  We show such a fit, with the very soft component modelled
by a MCDBB, in Table~\ref{complexfits}.  This fit is statistically
acceptable for the data, and an improvement over the single powerlaw
continuum model at 99.97\% ($> 3\sigma$) confidence.  Note, however,
that the addition of a classical blackbody emission model (with $kT =
0.1$ keV) rather than a MCDBB model provides an equally acceptable fit
to the data.  The powerlaw continuum slope is consistent (within the
large errors) in both epochs, though the inferred absorption column is
much lower in the absence of a very soft component in the second
observation.

\subsubsection{CXOU J123551.7+275604 (NGC 4559 X-1)}

This is the most luminous ULX in the sample, with its observed
luminosity at, or above, $10^{40} \ergsec$.  Its X-ray spectrum was
observed to soften between the observations, with the X-ray emission
in the second, more X-ray luminous epoch represented by a
significantly softer powerlaw photon index than the first ($\sim 2.16$
against $\sim 1.91$).  However, its X-ray spectrum is not well-fit by
either simple model, with the MCDBB model rejected at high statistical
significance.  Two-component models incorporating a very soft MCDBB
plus a powerlaw continuum offered improvements to the fit, as shown in
Table~\ref{complexfits}, though only at the 69\% and 94\% significance
levels in the first and second epochs respectively.  A classic
blackbody component offered a superior improvement in the second
epoch, at the 96\% significance level according to the F-statistic.

\begin{figure}
\centering
\includegraphics[width=5.2cm,angle=270]{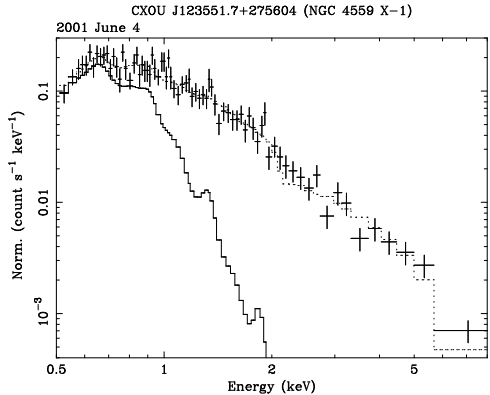}
\caption{The second epoch count rate spectrum of CXOU J123551.7+275604.  
The best-fit of the powerlaw plus {\small MEKAL} model to the
overall spectrum is shown as the dotted line with the contribution 
of the {\small MEKAL} component highlighted (solid line).}
\label{mekalplot}
\end{figure}

Curiously, a substantial improvement in the fit to the second epoch
data was obtained when the MCDBB component was replaced with a {\small
MEKAL} solar abundance thermal plasma with $kT \sim 0.18$ keV
(Table~\ref{complexfits}). For two additional free parameters, the
reduction in $\chi^2$ of 23.3 implies a 99.99\% significance level
(i.e. $> 3.5\sigma$) in terms of the F-test.  Figure~\ref{mekalplot}
shows the resulting best-fit spectrum and the contribution of the
{\small MEKAL} component. For comparison, a similar model (powerlaw
plus $kT \sim 0.18$ keV {\small MEKAL}) was fitted to the first-epoch
data and no significant improvement to the best-fit was obtained, with
the contribution of the putative thermal component limited to no more
than 3\% of the 0.5 -- 8 keV flux.

However, the MEKAL fit relies heavily on the fact that a blend of
lines between 0.5 -- 0.75 keV (predominantly OVII and OVIII), when
combined with a factor $\sim 2$ increase in the model $N_H$, gives a
good match to the additional soft emission apparent in the
second-epoch spectrum.  Is this thermal component real or simply an
artefact of fitting a fairly complex model to data with limited
spectral resolution and poor statistics (N.B. $\sim 600$ counts out of
$\sim 2000$ in the full spectrum originate from the {\small MEKAL}
component)?  Clearly there is no direct evidence in the second-epoch
spectrum for individual lines (though this is perhaps consistent with
low temperature of the plasma and the $> 100$ eV spectral resolution
of the ACIS-S detector below 1 keV). On the other hand when we allow
the abundance of O, Ne and Fe (which produce the most significant line
features in the $kT \sim 0.18$ keV plasma) to vary in the {\small
VMEKAL} model, we obtain a best-fit of 1.3 solar and a 90\%
lower-limit of 0.33 solar, implying that there is a strong preference
for a substantial contribution from emission lines.  {\it However,
this is on the basis of a spectral model which may not be correct}. We
conclude that an interpretation of the additional soft flux present in
the second-epoch in terms of the emergence of a bright thermal
component is potentially very interesting (see Section 5.3), albeit
highly speculative.

Finally, we note that as this paper was undergoing the refereeing
process these \chan spectra were published by Cropper et al. (2003) as
part of an \xmm study of this source.  They do not report the possible
detection of a {\small MEKAL} component, though they did not
explicitly look for it.  Instead they fit a particular model to the
\chan data, based on fits to the superior quality (and later epoch)
\xmm spectra, comprising a sub-solar abundance absorber (TBVARABS, set
at 0.31 times solar) applied to powerlaw plus blackbody emission
components.  We have applied the Cropper et al. model, plus variants
incorporating soft MCDBB and {\small MEKAL} components, to the
second-epoch spectrum and do get a marginal improvement in each case
with respect to our original fits ($\Delta \chi^2 \sim 2 - 4$);
nevertheless the powerlaw plus {\small MEKAL} model still clearly
provides the best result in terms of the minimum $\chi^2$.

\subsubsection{CXOU J123558.6+275742 (NGC 4559 X-4)}

This object was the only ULX in the sample with an X-ray spectrum
clearly best-fit with a MCDBB model.  The inferred inner accretion
disc temperatures were both greater than 1 keV, consistent with
previous {\it ASCA\/}, and some \chan observations of ULX well-fit by
this model (e.g. Colbert \& Mushotzky 1999; Makishima et al. 2000;
Roberts et al. 2002).  The luminosity of the ULX more than doubles in
the five months between observations, and this is associated with a
slight hardening of the X-ray spectrum, with the best-fit temperature
of the inner accretion disc rising from $\sim 1.1$ to $\sim 1.3$ keV.

\subsubsection{CXOU J132938.6+582506 (NGC 5204 X-1)}

The X-ray spectrum of this ULX was far better fit by a powerlaw
continuum model than a MCDBB in both epochs, with it constituting a
statistically-acceptable fit in the second epoch.  It was a poorer fit
in the first epoch, though this observation was affected by a pile-up
fraction of in excess of 10\% of all events (see above).  The {\small
XSPEC v.11.2} parameterisation of the Davis (2001) pile-up correction
model was used to correct for this effect, the results of which are
shown in Table~\ref{complexfits}.  This resulted in a much improved
fit to a powerlaw continuum spectrum, with the inferred photon index
softening considerably.  Hence, whilst the two uncorrected spectra
appear to have very different hardnesses, after the correction for
pile-up they are both revealed to be intrinsically very soft X-ray
spectra.  The photon indices of the powerlaw continuum fits are
consistent within the errors after pile-up correction, albeit with a
slightly harder value in the far more luminous first observation.

\subsection{Temporal properties}

\subsubsection{Short-term variability}

\begin{figure}
\centering
\includegraphics[width=5cm,angle=270]{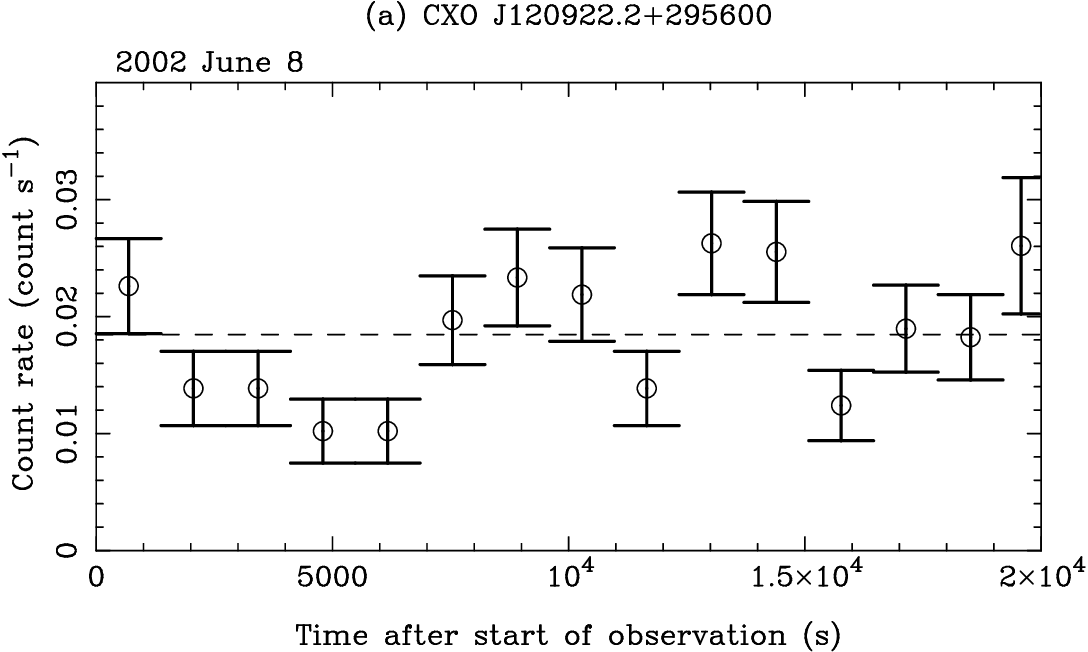}\vspace*{0.3cm} 
\includegraphics[width=5cm,angle=270]{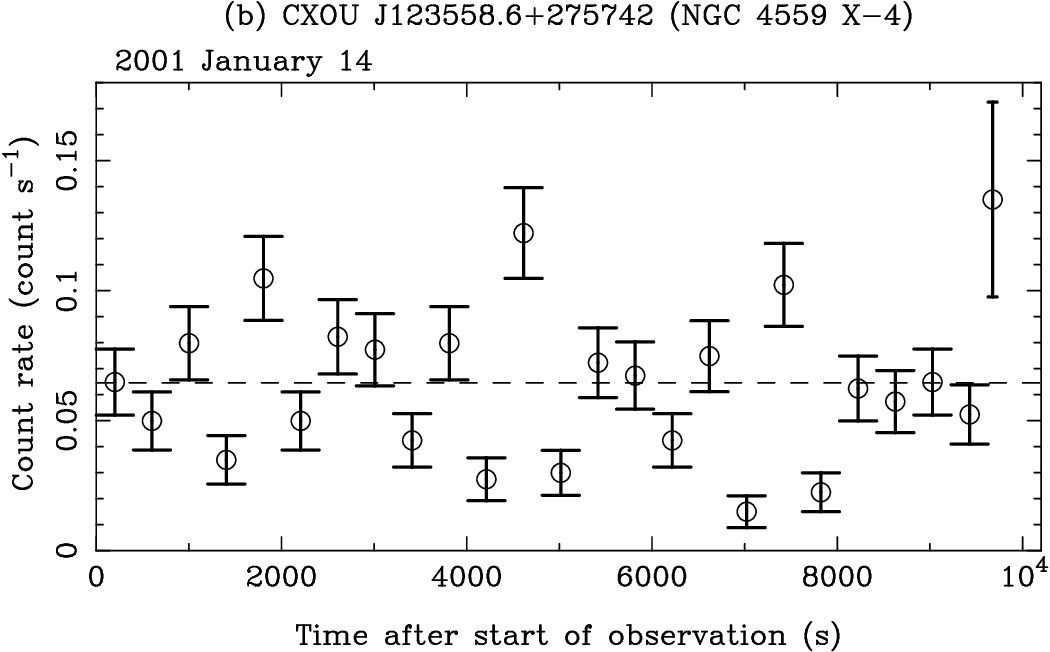}
\caption{The short-term variability observed in two ULXs.  The data in
both panels is binned such that, at the mean count rate (shown by the
dashed line) each temporal bin would contain 25 counts.}
\label{stvar}
\end{figure}

We derived short-term lightcurves for seven of the eight ULXs in both
of their observation epochs, using the {\small CIAO} routine {\small
LIGHTCURVE}.  The exception was CXOU J120922.6+295551, which was too
faint for this analysis with only 59 and 90 counts detected in total
per epoch.  Each lightcurve was binned to an average (over the
observation) of 25 counts per bin, giving temporal resolutions ranging
from $\sim 60$ seconds in the best case, to $\sim 2500$ seconds.  The
resulting data was tested for gross variability using a $\chi^2$ test
against the hypothesis of a constant count rate.  Five of the ULX
showed no strong evidence for short-term temporal variability in
either epoch, with a reduced-$\chi^2$ statistic of 1.25 or less.
However, two ULX showed strong evidence for short-term variability in
one of their two observation epochs.  These lightcurves are plotted in
Figure~\ref{stvar}.  The appropriate $\chi^2$/degrees of freedom
statistics are 39.5/14 for the 2002 June observation of CXO
J120922.2+295600 (which had a 1407 second resolution), and 190/24 for
CXOU J123558.6+275742 in its 2001 January observation (at 404 second
resolution).  The other observations of both sources were consistent
with a constant flux, implying that the short-term variability states
are themselves a transient phenomenon.  These highly-variable states
both occur when the ULX in question is (on average) at the lower of
its two observed X-ray fluxes.

We also performed a Kolmogorov-Smirnov test comparing the cumulative
photon arrival times with the expected arrival times, assuming each
source flux was invariant with time.  This provides a separate
indicator of short-term variability, free of any possible gross
binning effects, and sensitive to lower-amplitude (10 -- 20\%)
sustained changes in the ULX count rates than the $\chi^2$ test.  We
performed this test on all eight ULXs.  It confirmed that the 2002
June observation of CXO J120922.2+295600 was variable at $> 99\%$
probability.  Curiously, the 2001 January observation of CXOU
J123558.6+275742 was only variable at the $\sim 90\%$ level according
to the Kolmogorov-Smirnov test.  Again, none of the other ULX appeared
variable according to this test.

\begin{figure*}
\centering
\includegraphics[width=17.5cm]{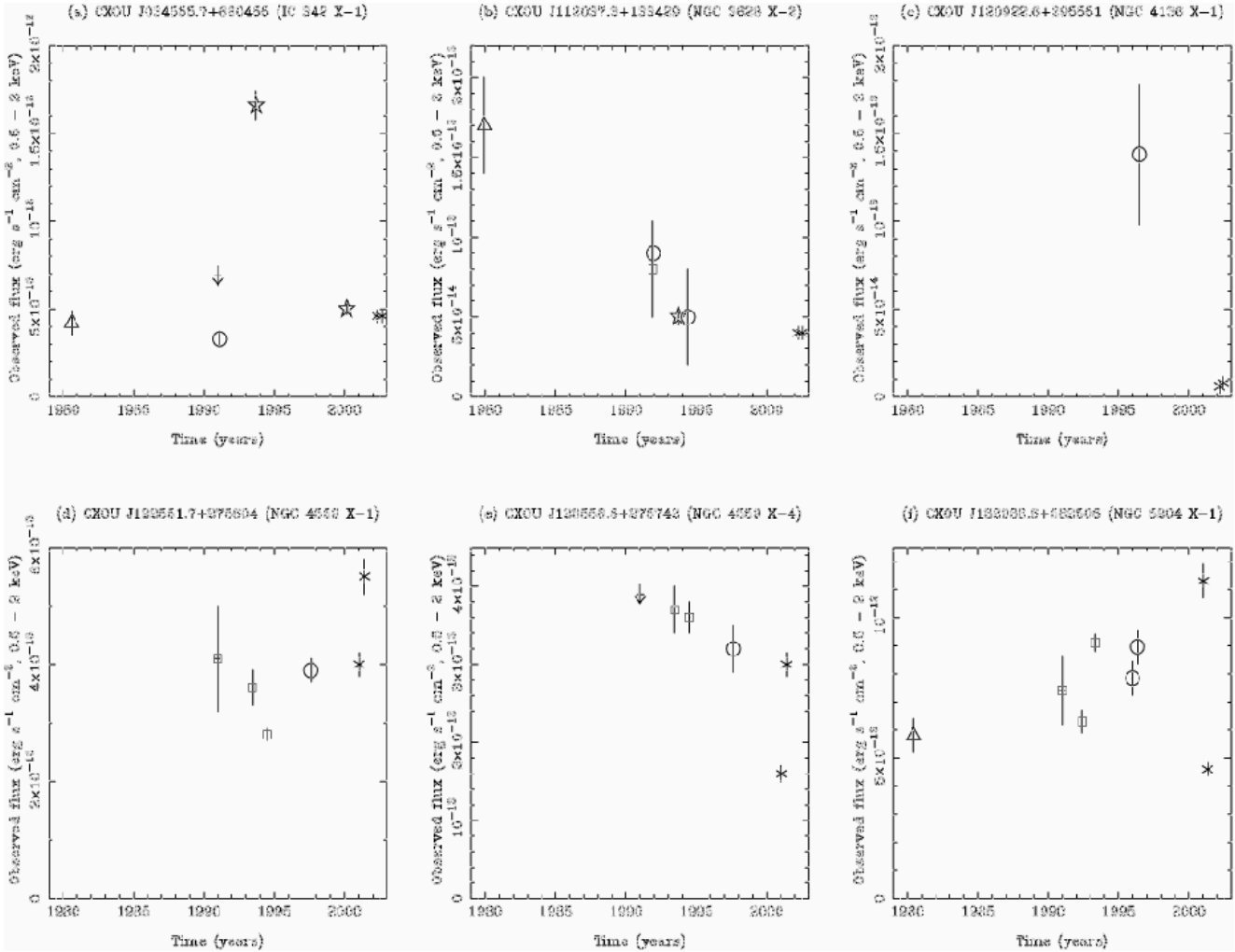}
\caption{Long-term lightcurves for six of the ULXs.  The data points
shown originate from \ein IPC (triangles), \ro PSPC (squares), \ro HRI
(circles), \asca (stars) and \chan (asterisks) measurements, which
have been converted to a 0.5 -- 2 keV flux as described in the text.
Each panel is normalised independently to highlight the variability of
each ULX.}
\label{ltvar}
\end{figure*}

In the January 2001 lightcurve of CXOU J123558.6+275742 there are four
``peaks'' in the X-ray emission, separated by regular intervals of 6
-- 7 bins, during which the X-ray flux is 4 -- 5 times brighter than
in the lowest flux bins.  This hints at a possible characteristic
timescale for the variability of $\sim 2500$ seconds for the peaks.
It is also noticeable that the lightcurve behaves similarly between
the peaks (this quasi-repetitive behaviour is the likely reason for
the lower probability of this source being variable according to the
Kolmogorov-Smirnov test).  Similar regular variability in the form of
``quasi-periodic flaring'' has recently been revealed in three
separate observations of an extremely variable ULX in M74, CXO
J013651.1+154547 (Krauss et al. 2003).  However, the characteristic
timescales of these variations are somewhat longer, inferred to be at
$\sim 3600 - 7900$ seconds, and the flares have a larger amplitude,
reaching more than an order of magnitude brighter than the quiescent
level for this ULX.  Claims of actual periodic variations have so far
only been made for three ULXs.  A period of $7620 \pm 500$ seconds was
reported for a bright state of the ULX M51 X-7 (Liu et al. 2003),
albeit on the basis of a 15 ks observation.  A 7.5 hr period was
derived for a ULX coincident with the Circinus galaxy (Bauer et
al. 2001), though given the proximity of Circinus to the Galactic
plane and the similarity of the periodicity to a long-period AM Her
system it is possible that this source is a foreground interloper.
The only other apparently periodic variations in an ULX were reported
by Sugiho et al. (2001) for IC 342 source 3, but these were very small
amplitude ($\sim 5\%$) and on a timescale of 30 -- 40 hours.

\subsubsection{Long-term variability}

Six of the eight ULXs have been detected and catalogued in the data
from previous X-ray observatories.  The long-term variability
lightcurves for these objects are plotted in Figure~\ref{ltvar}, using
data from the {\it Einstein\/}, \ro and \asca observatories in
addition to our new \chan data.  Each lightcurve is plotted in terms
of the observed 0.5 -- 2 keV flux of the ULX, with this particular
band chosen for its commonality between the missions.  The flux is
either directly measured from our best fit spectral models, in the
case of the \chan data, or is a flux quoted from previous spectral
measurements of the source in question (using \asca and some \ro PSPC
data: see Appendix A for references), or are calculated from the
observed count rates or fluxes given in a wider band.  In the latter
cases, we assume spectral models that are coarse averages of the two
\chan observations (e.g. an absorbed powerlaw continuum model with
$\Gamma = 2$ and \nh $= 2 \times 10^{20} \atpcm$ for CXOU
J123551.7+275604), and use either {\small PIMMS} or {\small XSPEC} to
normalise the observed count rate into an observed 0.5 -- 2 keV flux.
Finally, we correct the \ein and \asca count rates of CXOU
J112037.3+133429 for confusion with the two moderately-bright sources
immediately to its north-northwest (c.f. Figure~\ref{ulxlocations}),
and we similarly correct the \ro PSPC data for CXOU J123558.6+275742
for confusion with other sources in proximity to the nucleus of NGC
4559, both on the basis of the \chan observations.

The long-term lightcurves of the ULXs are obviously very poorly
sampled, with between three and eight observations, each typically
only a few hours long, covering a baseline from 6 to 23 years.
However, despite the low quality, there does appear to be some variety
evident in the behaviour of the different ULX.  Two sources - CXOU
J123551.7+275604 and CXOU J132938.6+582506 - appear to have varied
quite randomly over their observational histories, with a maximum
change in their observed flux of little more than a factor of two.  On
the other hand, the average flux of CXOU J112037.3+133429 appears to
have consistently decayed over 23 years of observations to $\sim 25\%$
of the value inferred from the \ein IPC.  A possible decay is also
observed from CXOU J123558.6+275742, with the exception of a large dip
in its first \chan epoch, which coincides with the epoch of extreme
short-term variability in this ULX.  A further source whose flux has
decayed since its initial detection is CXOU J120922.6+295551, which
has faded by a factor $\sim 20$ over the six years between its initial
detection in a \ro HRI image and the current \chan observations.
Finally, CXOU J034555.7+680455 appears at around the same 0.5 -- 2 keV
flux ($4 \pm 1 \times 10^{-13} \ergcms$) in all but one of its
observations.  The exception is the 1993 September observation in
which it displayed a high/soft spectral state, with an average flux
four times higher than in other epochs.  This may indicate that the
high/soft state is rare in this source.

We do not present long-term lightcurves for the two previously
undetected ULXs.  The 0.5 -- 2 keV flux of CXOU J120922.2+295600 in
the \chan observations was $\sim 5 - 7 \times 10^{-14} \ergcms$; if
this flux was the same during the previous \ro HRI observation of this
galaxy, it would have remained undetected (NGC 4136 X-1 $\equiv$ CXOU
J120922.6+295551 was a marginal detection at $\sim 1.5 \times 10^{-13}
\ergcms$).  On the other hand, CXOU J123557.8+275807 was observed to
have a flux of $3 - 6 \times 10^{-14} \ergcms$ in the \chan
observations, which compares to detection limits of around 1 and $4
\times 10^{-14} \ergcms$ in the \ro PSPC and HRI observations of NGC
4559 respectively.  This implies that this ULX is possibly a transient
source.

Finally, our \chan observations reveal that two out of the five target
ULXs have faded to X-ray luminosities that are below the arbitrary
limit of $10^{39} \ergsec$ that we use to demarcate the ULX population
from the ``normal'' X-ray source population in nearby galaxies.  An
important aim for future studies will be to determine how common this
behaviour is for ULXs, and in particular to define their duty cycle,
allowing us to estimate the true number of potential ULXs in nearby
galaxies.

\section{Discussion: clues to the nature of ULXs from their behaviour?}

\subsection{Similar behaviour}

Are there any general trends revealed in our data that can help cast
light on the nature of ULXs?  Several similarities across the sample
are highlighted by the analysis in the previous sections.  They are
all spatially consistent with point-like sources, and when their
ubiquitous long-term variability (and in some cases extreme short-term
variability) is considered it is very likely that their X-ray emission
originates in a single X-ray emitting system.  The majority of the
best-fit X-ray spectral models show an absorption column well in
excess of that along the line-of-sight to each ULX through our own
galaxy.  This may either be an additional column through the host
galaxy, or perhaps material intrinsic to the ULX itself.  The lack of
variation in the absorption column where it is constrained by the same
physical model at both epochs (c.f. Table~\ref{xspecfits1}) suggests
that it may not originate in the very close proximity of the ULX,
although it may still be associated with gas and dust in their
immediate environment.

It is likely that the majority of the ULXs are associated with the
young stellar populations of the active star formation regions in
their host galaxies.  The issue of whether this implies that they do
not contain an IMBH primary remains controversial.  Models for IMBH
formation in young, dense stellar clusters (e.g. Ebisuzaki et
al. 2001; Portegies-Zwart \& McMillan 2002) certainly argue that star
formation regions could host IMBHs, though the observed displacement
between stellar clusters and ULX positions in the Antennae (Zezas et
al. 2002) argues that at least in this case many of these ULXs are
runaway binaries (i.e. ordinary X-ray binaries kicked out of the
stellar clusters by the supernova explosions that formed the compact
primary).  If alternately the IMBHs were formed before the current
epoch of star formation, for instance as the remnants of the
primordial Population III stars (Madau \& Rees 2001), it is unlikely
that they could accrete sufficient material to appear ULX-bright even
in the gas-rich environment of star-formation regions (c.f. Miller \&
Hamilton 2002), therefore they are reliant upon capturing a secondary
star to accrete from.  The chances of this occuring could be higher
within the dense stellar environments of star forming regions.
However, whatever the origin of the IMBH, King (2003b) argues that
extremely large underlying populations of these objects are required
in order to account for the numbers of ULXs observed in the brightest
starburst galaxies such as the Cartwheel galaxy (c.f. Gao et
al. 2003).  This strongly argues that ULXs associated with star
formation are, as a class, dominated by ``ordinary'' HMXBs that
somehow exceed their Eddington limit.  The likely association of the
ULXs in our sample with young stellar populations could simply be the
extension of the ULX -- star formation relationship into lower
star-formation rate ``normal'' spiral galaxies.  If so, then it is
likely that our objects may also be predominantly HMXBs.

Next, \chan does not detect extreme short-term X-ray variability (on
the scale of hundreds of seconds over a baseline of three to six
hours) as a common feature of these ULXs - it is only present in two
of 14 observations - implying that in the majority of ULX states
either the X-ray emission is reasonably constant, or it varies on much
shorter timescales than we have sampled.  The absence of extreme
short-term variability, on the basis of \chan observations, has also
been noted for many other ULXs in spiral galaxies (c.f. Strickland et
al. 2001; Roberts et al. 2002), The lack of this behaviour in the
majority of ULXs argues strongly against the relativistic beaming
model as the dominant emission mechanism for ULXs, since presumably
this would require a remarkably stable jet to avoid large amplitude
flux variation, due to both its narrow beam angle and the strong
amplification of any variations via Doppler boosting.

On the other hand, the absence of extreme long-term variability is
also interesting, with most of the ULXs seen to vary only by factors
of up to $\sim 4$ over periods of up to twenty years.  This suggests
that they are all persistent X-ray sources, rather than short
duty-cycle transients.  Interestingly, a comparison with the 18
dynamically-confirmed Galactic black hole binaries shows that the only
persistent sources over the last $\sim 30$ years are the three HMXBs
(McClintock \& Remillard 2003).  Furthermore, observations of one of
the HMXBs, Cyg X-1, with RXTE have shown its flux to vary by no more
than a factor of $\sim 4$ over seven years of continuous observations
(Pottschmidt et al. 2003).  These similarities may be another clue
linking ULXs to HMXBs, although whether this is a fair comparison
given that the three Galactic HMXB/BH sources are all wind accretors
with L$_{\rm X}$ well below $10^{39} \ergsec$ is open to debate.

A further trend revealed by the data analysis is that the X-ray
spectra of the ULXs are generally better fit by powerlaw continua than
by MCDBB models, with only one out of six ULXs showing a preference
for the latter (though we note that in three of the ten powerlaw
spectra a combination of low signal-to-noise and a low $\chi^2$/dof
value implies that we cannot exclude MCDBB models to a high
statistical significance).  This appears to contradict previous
results, particularly those reported from \asca observations by
Makishima et al. (2000), which focussed on ULXs well fit by a MCDBB as
possible examples of the high/soft state for accreting 10 - 100
M$_{\odot}$ black holes.  So if these ULXs are comparable to Galactic
sources, but are not in the high/soft state, what state are they in?
To answer this question we refer to the recent work by McClintock \&
Remillard (2003), which provides new definitions of the classic black
hole states, with firm observational diagnostics based on the results
of more than six years of observations with RXTE.  However, since we
have neither the photon statistics in the \chan data to derive a power
density spectrum, or any information on the X-ray spectrum beyond a
maximum of 10 keV, we are reliant solely on the photon indices of the
powerlaw continua to provide a state diagnostic.  The two states
described by McClintock \& Remillard (2003) that provide the best
candidates for the powerlaw-dominated spectra are the low/hard state
(which is related to the emission of radio jets) and the steep
powerlaw state (previously known as the very high state), with the
distinction in photon indices being that the low/hard state typically
has values of $\Gamma \sim 1.5 - 2$, whereas the steep powerlaw state
has values of $\Gamma > 2.4$.  Using this distinction, it is clear
that only one source (CXOU J132938.6+582506) has a sufficiently steep
powerlaw to be in the latter class ($\Gamma \sim 2.8 - 3.0$), with all
the other sources more consistent with the low/hard state.  It is also
notable that Galactic black holes in the steep powerlaw state tend to
have a substantial contribution from a MCDBB component with $kT_{in}
\sim 1 - 2$ keV; we do not see any such composite warm MCDBB +
powerlaw continuum spectrum (although the \chan bandpass may not be
the optimum choice for recognising such a spectrum).

This provides an observational argument against the suggestion of
Kubota, Done \& Makishima (2002; see also Terashima \& Wilson 2003 for
further discussion) that the the powerlaw spectral state of ULXs must
be the very high state.  Instead, the slopes of the four out of five
ULXs best described by powerlaw spectra suggest that the sources are
in the low/hard state.  Until recently, this would have suggested that
the ULX was accreting at a relatively low fraction (maybe $\sim 5\%$)
of the Eddington luminosity of the compact object, as black hole
states were simplistically assumed to be a function of mass accretion
rate (see e.g. Esin, McClintock \& Narayan 1997).  This would imply
black hole masses of 1 -- 2 $\times 10^3$ M$_{\odot}$ for these ULXs,
i.e. IMBHs.  However, it is now clear that the black hole state is not
a direct function of mass accretion rate (Homan et al. 2001).  This
suggests that we cannot rule out observing the low/hard state at or
near the Eddington limit for a stellar-mass black hole.  However, it
is still true that in most known cases the low/hard state occurs well
below the Eddington limit, with the one exception being GRS 1915+105
which reaches luminosities $> 10^{38} \ergsec$ in this state
(McClintock \& Remillard 2003).  For ULXs to be low/hard state
stellar-mass black hole binaries we would obviously have to rely on
some means of boosting the apparent luminosity to explain their
super-Eddington fluxes, e.g. anisotropic emission.  However,
observations of the low/hard state in Galactic black holes have
suggested that the accretion disc could be truncated at large
distances from the black hole ($\ga 100 R_g$), which would present a
challenge for models in which the anisotropic emission originates in
``funneling'' of the radiation by a geometrically-thick inner
accretion disc (c.f. King 2003a) (though see McClintock \& Remillard
2003 for a discussion of whether the inner disc is truly truncated in
the low/hard state).  This suggests that the easiest interpretation of
this state is that we are observing the low/hard state from accretion
onto an IMBH.

The arguments presented above assume that direct analogies to Galactic
black hole systems are appropriate for ULXs, when the majority of
Galactic black holes spend the majority of their time emitting at far
below the $10^{39} \ergsec$ threshold of ULXs.  It is possible then
that the hard powerlaw state we are seeing is actually unique to the
very high accretion rates required if ULXs are truly the extreme high
luminosity end of the stellar-mass black hole population.  If this is
the case, then super-Eddington ratios (i.e. \lx/L$_{\rm Edd}$) of up
to 10 are required for the ULXs discussed in this paper; this is
consistent with the predictions of Begelman (2002) for true
super-Eddington discs, and within a factor two of the ratio predicted
from simple geometric funelling by Misra \& Sriram (2003).  Of course,
we cannot rule out accretion onto IMBHs as an explanation for the
spectra of any of our sources, but it is worth noting that it is now
becoming clear that Galactic black holes do show epochs of apparent
super-Eddington emission (McClintock \& Remillard 2003).  Perhaps the
most pertinent example of these Galactic sources is GRS 1915+105,
which spends a significant fraction of its time emitting at
super-Eddington rates for its $\sim 14$ M$_{\odot}$ black hole (Done
et al. 2003), and has been in outburst continually for the past 11
years.  Hence if viewed by an observer outside our galaxy, GRS
1915+105 could frequently appear as a ULX.

\subsection{Contrasting behaviour}

A remarkable feature of the analysis in the previous sections is the
diversity in the observed properties of the ULXs, with the differences
between the objects being more pronounced than any similarities.  To
obtain a deeper understanding of these objects, we adopt a more
complete view than simply considering each characteristic in
isolation.  We approach this by compiling a summary of their
observational characteristics in Table~\ref{srcdiags}.

\begin{table*}
\caption{A summary of the ULX characteristics.}
\begin{tabular}{lcccccl}\hline
ULX (CXOU J)	& Epoch	& \lx$^a$	&
\multicolumn{2}{c}{Variability}	& Best-fit spectral model$^b$	&
Environment \\ 
 & & & Short-term	& Long-term	&
\\\hline
034555.7+680455	& 2002-04-29	& 4.4(5.9)	& No	& Hard state +
soft flare	& PL 	& Spiral arm \\
 & 2002-08-26	& 4.4(6.4)	& No 	& & \\
112037.3+133429	& 2002-04-06	& 0.7(0.8)	& No	& Steady flux
decay	& PL 	& Outer disc \\
 & 2002-07-04	& 0.6(0.7)	& No	& & \\
120922.2+295600	& 2002-03-07	& 2.3(2.6)	& No	& [N/A]	& PL + 
absorbed MCDBB	& Spiral arm \\
 & 2002-06-08	& 1.7(1.9)	& {\bf Yes}	& & $\rightarrow$ PL \\
120922.6+295551	& 2002-03-07	& 0.2(0.2)	& No	& Large drop
& [N/A] 	& Spiral arm \\
 & 2002-06-08	& 0.2(0.2)	& No	&  & \\
123551.7+275604	& 2001-01-14	& 9.8(10.0)	& No	& Random
& PL (+ soft component?) 	& Faint outer spiral arm \\
 & 2001-06-04	& 11.6(12.5)	& No	& & $\rightarrow$ PL + MEKAL \\
123557.8+275807	& 2001-01-14	& 1.4(1.5)	& No	& Transient?
& [N/A] 	& Inner disc \\
 & 2001-06-04	& 2.6(2.9)	& No	& & \\
123558.6+275742	& 2001-01-14	& 4.2(4.3)	& {\bf Yes}	&
Gradual fade or random?	& MCDBB 	& Inner disc/bulge \\
 & 2001-06-04	& 8.9(9.1)	& No	& & \\
132938.6+582506	& 2001-01-09	& 5.7(6.9)	& No	& Random
& Soft PL 	& Inner disc \\
 & 2001-05-02	& 1.8(2.4)	& No	& & \\\hline
\end{tabular}
\begin{tabular}{l}
Notes: $^a$ Observed 0.5 -- 8 keV luminosity in units of $10^{39}
\ergsec$.  Figures in parentheses are the intrinsic values.  The
luminosity of\\
CXOU J120922.6+295551 is converted from the count rate assuming a
powerlaw continuum with $\Gamma = 1.8$ and foreground absorption,\\
whereas the luminosities for CXOU J123557.8+275807 are derived from a
rough spectral fit to its second epoch data (an absorbed MCDBB\\ with
\nh~ $\sim 9 \times 10^{20} \atpcm$ and $kT_{in} \sim 1.2$ keV).
$^b$ We use PL as shorthand for a powerlaw continuum.  [N/A]
indicates\\ that we did not have sufficient quality data for the
analysis.
\end{tabular}
\label{srcdiags}
\end{table*}

One immediate curiosity highlighted by Table~\ref{srcdiags} is that
the two sources that display short-term variability appear very
different.  Whilst both the sources are at the lower of their two
observed luminosities when variable, their average X-ray spectra
during this epoch contrast greatly, with CXOU J120922.2+295600
displaying a hard powerlaw continuum ($\Gamma \sim 1.55$) whereas CXOU
J123558.6+275742 has a MCDBB form.  The short-term variability and
powerlaw spectral form of CXOU J120922.2+295600 may be consistent with
its X-ray emission in this state being dominated by a variable,
possibly relativistically-beamed jet.  However, the thermal accretion
disc spectrum of CXOU J123558.6+275742 argues against this
interpretation.  Further studies of short-term variation in ULXs,
based on higher signal-to-noise ratio data, have suggested a number of
alternate physical mechanisms including rapid variations in the ULX
accretion rate (La Parola et al. 2003), magnetic reconnection events
in the accretion disc corona (Roberts \& Colbert 2003) and an
optically thick outflow (Mukai et al. 2003).

\begin{figure}
\centering
\includegraphics[width=6.5cm,angle=270]{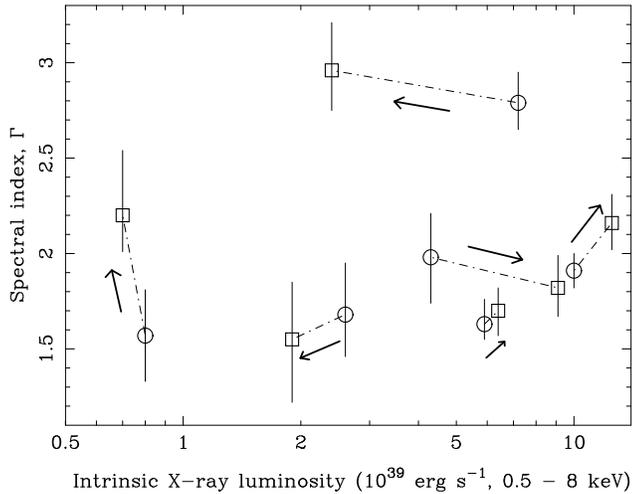}
\caption{Spectral variability in the ULX as a function of intrinsic
X-ray luminosity.  The first epoch \chan data is indicated by an open
circle, and the second by a square, with the two observations of each
source joined by a dot-dash line.  The evolution of the photon index
with luminosity is highlighted by the arrows, and the errors are the
90\% confidence intervals for the spectral parameters.}
\label{specvar}
\end{figure}

A related issue is the long-term variations in the X-ray spectra of
these sources revealed by the \chan data.  We note, with reference to
Table~\ref{xspecfits1}, that all the spectra appear to show changes
between the two observational epochs.  We summarise these changes in
Figure~\ref{specvar} by plotting the best-fit powerlaw photon index
for each dataset against the derived intrinsic 0.5 -- 8 keV X-ray
luminosity for that observation.  Though Figure~\ref{specvar}
highlights that the 90\% confidence intervals for the photon index
measurements overlap in almost all cases, consistent with little or no
significant change in the photon indices, the best fit values do
appear to show some level of variation.  The data suggest that three
of the ULX spectra are softer in a higher luminosity state, and three
are harder.  It has been noted elsewhere (e.g. Kubota et al. 2001; La
Parola et al. 2001) that some ULX undergo a low/hard to high/soft
state transition, reminiscent of the behaviour of classic Galactic
black hole X-ray binary systems.  Other recent observations have
highlighted the opposite behaviour, i.e. a high state in which the
spectrum is harder, for instance in M51 X-7 (Liu et al. 2002) and in
four out of five ULXs in the Antennae (Fabbiano et al. 2003), one
interpretation being that the accretion disc heats up (and hence
spectrally hardens) as the luminosity of the system increases.  The
equal split between ``hardening'' and ``softening'' ULXs in our sample
suggests that both transitions are common, implying either the ULX
population is heterogeneous, with at least two flavours of accretors
behaving differently, or that those sources composing the bulk of the
ULX population are capable of displaying multiple states and modes of
transition between those states.

\subsection{(Dis)appearing spectral components}

Table~\ref{srcdiags} also highlights a more spectacular flavour of
spectral variation: components that appear to be present at one epoch
but not during the other.  This occurs in two of the ULXs.  In CXOU
J120922.2+295600 we see a very soft component, fit by a MCDBB model
with $kT_{in} = 0.12$ keV, that is present (though only at $3\sigma$
confidence) in the more luminous 2002 March observation but not in
2002 June.  A similar soft ($kT < 0.2$ keV) MCDBB component has been
reported in other ULX, notably from \xmm data for two ULXs in NGC 1313
(Miller et al. 2003), where it is interpreted as spectroscopic
evidence for the ``cool'' accretion disc expected to be present around
an IMBH.  In CXOU J120922.2+295600 this component disappears in the
three-month period between observations.  This could be analogous with
a classic high/soft to low/hard state transition as seen in Galactic
black-hole X-ray binary candidates, but potentially involving an IMBH
system in this case.  However, there are other potential explanations.
Perhaps the best alternative is that the soft component is originating
in an optically-thick outflowing photosphere around a black hole X-ray
binary, as suggested by Mukai et al. (2003) to account for a highly
luminous ultra-soft ULX in M101 (a phenomenon also described as a
``black hole wind'', see King \& Pounds 2003; Fabbiano et al. 2003).
Such a component would have a blackbody spectrum with a temperature
$kT \sim 0.1$ keV, consistent with the first epoch observation
(c.f. section 4.2.3), and might easily switch off between the two
observations of the ULX.  However, there are arguments against the
soft component originating in a black hole wind, most notably in the
case of sources that also have a strong hard powerlaw continuum
component in their spectrum (e.g. M81 X-9; Miller, Fabian \& Miller
2003).  In this case the powerlaw component must originate in shocks
outside the photosphere, but in order to have comparable luminosity to
the soft component Miller et al. argue that the photospheric radius
must actually be of the order of the Schwartzschild radius, implying
no photosphere for these sources, and hence that the soft component
must originate in the accretion disc.

The other ULX to show a changing spectral form was CXOU
J123551.7+275604.  In this case the spectrum showed a marked
softening between the two \chan observations. Earlier we noted that
the additional soft flux present in the second-epoch could be modelled 
in terms of the emergence of soft ($kT =0.18$ keV) optically-thin 
thermal emission in the ULX spectrum, which is not present 
five months earlier.  Since individual line features are not evident 
in the raw data, this interpretion is very model dependent and 
needs to be checked via future observations. Bearing in mind these 
caveats we can, nevertheless, speculate as to how such thermal emission 
might be produced. 

The putative {\small MEKAL} component is emitting an incredible 
$\sim 2.7 \times 10^{39} \ergsec$ of 0.5 --8 keV X-ray luminosity
if the emission is isotropic.  To place this is context, this
an order of magnitude greater than the integrated diffuse luminosity
of several small nearby galaxies (c.f. Read, Ponman \& Strickland
1997).  We note that CXOU J123551.7+275604 might not be an unique
system; two other ULXs have been reported with possible thermal plasma
components, with a similar very soft component reported on the basis
of a joint \ro PSPC/\asca spectral fit of the Holmberg II ULX by
Miyaji, Lehmann \& Hasinger (2001), and a much hotter and
time-variable thermal component detected for an ULX in M51 (Terashima
\& Wilson 2003).  

What processes might produce thin-thermal plasma emission in the X-ray
spectrum of a ULX?  Miyaji, Lehmann \& Hasinger (2001) suggest the
presence of one or more supernovae in the immediate environment of the
Holmberg II ULX, but we can reject this for CXOU J123551.7+275604
given the rapid appearance of the plasma and the lack of a reported
recent supernova at this position.  In contrast, Terashima \& Wilson
(2003) draw an analogy to the X-ray spectrum of some HMXBs in eclipse,
which are dominated by emission-lines that originate in
photoionization of the stellar wind by the hard X-ray emission of the
accreting primary (e.g. Cyg X-3; c.f. Liedahl \& Paerels 1996).
However, we cannot be observing thermal emission due to a reduction in
contrast with the direct emission from the ULX, since we measure a
higher overall flux in the second epoch observation when the plasma
component is present.  In any event, the plasma component we measure
appears far too soft, at 0.18 keV, to be directly analogous to the
photoionized stellar wind in sources such as Cyg X-3, which are
typically identified by strong emission lines at energies well above 1
keV.

The emission measure of the plasma is $n_e^2V = 2.9 \times 10^{63}$
cm$^{-3}$, where $n_e$ is the electron density in atom cm$^{-3}$ and
$V$ is the volume of the material.  Since the plasma is optically
thin, it must satisfy the limit $n_er < {{1}\over{\sigma_T}}$, where
$r$ is the pathlength through the plasma and $\sigma_T$ is the
Thompson scattering cross-section.  Combining these two estimates
gives $r_{min} \approx 3 \times 10^9$ km, implying that the plasma
occupies a region far larger than the black hole accretion disc.
However, as we argue above, it is probably too soft to be a
photoionized stellar wind, which leaves the alternative possibility
that it is a collisionally-ionized system.  This could arise if an
outflow from the ULX (perhaps in the form of a relativistic
jet\footnote{It is quite reasonable for an ULX to possess a
relativistic jet regardless of whether or not it dominates the X-ray
emission.  By way of analogy, Galactic black hole systems such as GRS
1915+105 (i.e. microquasars) certainly do possess relativistic jets
that do not dominate their X-ray emission.}) impacts upon a cloud of
material relatively close to the ULX.

We further speculate that this scenario might occur if the secondary
star in the system is an evolved high mass star, since these are known
to eject discrete shells of material, for example with luminous blue
variable (LBV) stars thought to lose up to $5 \times 10^{-3}$
M$_{\odot}$ yr$^{-1}$ in such discrete outburst events (van der Sluys
\& Lamers 2003 and references therein).  For the derived value of
$r_{min}$ we calculate an electron density $n_{e,max} \approx 4.9
\times 10^9$ cm$^{-3}$ and hence a mass of $4.8 \times 10^{-4}$
M$_{\odot}$ for the plasma, well within the ejection masses quoted
above.  To energize this mass of material to a temperature of $kT =
0.18$ keV requires an energy input of $E = 3N_ekT$, where $N_e$ is the
total number of electrons in the plasma, hence a total of at least
$\sim 4.8 \times 10^{44}$ erg is required.  In the 141 days between
the observations, this would require an average mechanical energy
input from the outflow of at least $\sim 4 \times 10^{37} \ergsec$.
We note that this energy input could be easily achievable by
relativistic jets; for example the W50 radio bubble surrounding SS 433
probably originates in inflation by the mildly-relativistic jets of SS
433, with an average mechanical energy input from the jets of $3
\times 10^{39} \ergsec$ (Dubner et al. 1998).  Hence this scenario
appears physically plausible.  

Finally, we note that if the collisionally-excited material does
originate in the ejection of matter from an evolved high-mass star,
then this scenario is consistent with the suggestion of King et
al. (2001) that ULXs are fed by thermal-timescale mass transfer from
an evolved secondary star in a HMXB.  However, it is not clear whether
this system could be an ``ordinary'' HMXB containing a stellar-mass BH
in accordance with the King et al. (2001) model, as Cropper et
al. (2003) conclude that its observational characteristics point
towards it containing an IMBH.  

Clearly future monitoring of the behaviour of CXOU J123551.7+275604
is required to resolve many the uncertainties as to its nature and
in particular to investigate whether outflow/stellar material collision
events occur in reality.

\section{Conclusions}

In this paper we have studied dual-epoch observations of five nearby
ULXs obtained with the \chan ACIS-S detector.  These have revealed
heterogeneity in the observed properties of ULXs, from the
characteristics of individual sources varying between the observation
epochs, to a wide range of observed X-ray properties across the
sample.  The main outcomes of our analysis are:

\begin{itemize}
\item
We detect all five target ULXs, though two have faded to luminosities
below $10^{39} \ergsec$, plus we find three additional ULXs within the
regions of the galaxies that we observe.
\item
The ULXs are all likely to be associated with the young stellar
populations residing in and around the star forming regions of their
host galaxies.  This is consistent with the discovery of large ULX
populations in starburst galaxies, extending the ULX -- young stellar
population link to ``normal'' spiral galaxies.  We estimate that no
more than two out of the eight ULXs originate in older stellar
populations on the basis of the ULXs per unit optical luminosity
measured in elliptical galaxies by Colbert \& Ptak (2002).
\item
All the ULXs are point-like at the 0.5-arcsecond on-axis \chan
resolution.
\item
We derive the spectral parameters for six ULXs in both observational
epochs, finding that the majority (five) are best fit by a powerlaw
continuum rather than a MCDBB model, contrary to some previous reports
for ULXs (e.g. Makishima et al. 2000).  Two of the five best fit by
powerlaws require an additional, very soft spectral component in at
least one observation.
\item
Extreme short-term variability (on a timescale of hundreds of seconds)
is observed in only two ULXs, and in these sources it is only seen in
one of two observations.  However, all the ULXs with archival data are
variable by factors of a 2 -- 4, over timescales of years.
\item
The lack of short-term variability, unless it occurs on timescales
much less than our temporal sampling, may indicate that the X-ray
emission of ULXs as a class is not dominated by
relativistically-beamed jets.
\item
By utilising the new black hole state diagnostics of McClintock \&
Remillard (2003), we see that four out of the five powerlaw-dominated
spectra have photon indices consistent with the low/hard state.  This
challenges the recent assertions that powerlaw-dominated spectra in
ULXs originate in the very high state.  However, if this truly is the
low/hard state it could pose problems for interpretations of ULXs
involving anisotropic stellar-mass black holes, and the simplest
explanation would be that we are seeing a low/hard state from
accretion onto an IMBH.  The alternative is that we might be observing
a state unique to the very high accretion rates required for
stellar-mass black hole models.
\item
We find that equal numbers of ULXs (three apiece) show spectral
``hardening'' and ``softening'' with increasing luminosity.  This
indicates either an underlying physical heterogeneity in the ULX
population, or perhaps that the bulk of the ULX population is composed
of sources that can behave in both ways.
\item
The X-ray spectrum of CXOU J123551.7+275604 is particularly
interesting, in that it is possible that a very soft and luminous
{\small MEKAL} component appears in the five months between the two
observations.  We speculate that this plasma (if real) originates in
the impact of an outflow (possibly relativistic jets) from the ULX on
material in its close vicinity.  One possible source of this material
could be in discrete mass ejection events from an evolved high-mass
secondary star.  The presence of an evolved high-mass companion star
is consistent with the scenario suggested by King et al. (2001), where
many ULXs are ordinary HMXBs with an evolved high-mass secondary,
though conversely this ULX is a very good candidate for an IMBH.
\end{itemize}

The properties outlined above portray ULXs as a class with very
heterogeneous properties.  Whilst some trends are emerging, such as
the likely relation between many ULXs and young stellar populations,
the lack of observable short-term variability, and a preference for
powerlaw continuum spectra, none of these provide conclusive
observational evidence to distinguish between the competing physical
models.  It could be that much of the confusion is because ULXs are a
physically heterogeneous class, including both stellar-mass black
holes, many of which could be in HMXBs, and IMBHs.  Perhaps, then, the
question we should be addressing is: to what degree are the ULX
populations heterogeneous?  An answer to this question is reliant upon
future large samples of high-quality X-ray data, plus detailed
multi-wavelength follow-up, to study these objects in much greater
detail.  It appears that we are still at the beginning of long road
towards understanding these extraordinary X-ray sources.

\vspace{0.2cm}

{\noindent \bf ACKNOWLEDGMENTS}

The authors would like to thank the anonymous referee for their
comments, which have greatly improved the interpretation of our
results.  We would also like to thank Paulina Lira for an initial
discussion on targets for this programme.  TPR gratefully acknowledges
support from PPARC.  This research has made use of the NASA/IPAC
Extragalactic Database (NED) which is operated by the Jet Propulsion
Laboratory, California Institute of Technology, under contract with
the National Aeronautics and Space Administration.  This research has
also made use of data obtained from the Leicester Database and Archive
Service at the Department of Physics and Astronomy, Leicester
University, UK.  The second digitized sky survey was produced by the
Space Telescope Science Institute, under Contract No. NAS 5-26555 with
the National Aeronautics and Space Administration.

\appendix

\section{Previous observations of the ULXs}

\subsection{IC 342 X-1 (CXOU J034555.7+680455)}

This ULX was first detected in an \ein IPC observation of IC 342, with
a 0.2 -- 4 keV X-ray luminosity of $3 \times 10^{39} \ergsec$
(Fabbiano \& Trinchieri 1987).  A subsequent \ro HRI observation
detected IC 342 X-1 in a similarly luminous state (Bregman, Cox \&
Tomisaka 1993; RW2000).  An early \asca observation showed IC 342 X-1
to be in a very luminous, highly variable state, in which its average
0.5 -- 10 keV luminosity surpassed $10^{40} \ergsec$ and it displayed
large amplitude variability on $\sim 1000$ s timescales (Okada et
al. 1997).  In this state its X-ray spectrum was well-fit by the
multi-colour disc blackbody model describing an accretion disc around
a black hole in a high/soft state (Makishima et al. 2000).  However, a
follow-up observation obtained late in the \asca mission showed the
ULX to have dimmed to a luminosity of $6 \times 10^{39} \ergsec$ and
undergone a spectral transition to a low/hard state, similar to the
behaviour observed in many Galactic and Magellanic black hole X-ray
binary systems (Kubota et al. 2001; Mizuno, Kubota \& Makishima 2001).
However, the low/hard spectral state has recently been re-interpreted
as a high/anomalous state where the X-ray spectrum is described by a
strongly Comptonised optically thick accretion disc, as observed in
many Galactic black-hole X-ray binaries experiencing a high mass
accretion rate (Kubota, Done \& Makishima 2002).  Finally, recent
William Herschel Telescope integral field unit observations have
revealed that this ULX lies at the centre of a large nebula, probably
a supernova remnant shell (Roberts et al. 2003 and references
therein).  The inferred initial energy input to the SNR is consistent
with a hypernova event, and highly excited [O{\small III}]
emission-line regions on the inner edge of the shell suggest that the
ULX is photoionizing its inner regions to produce an X-ray ionized
nebula.

\subsection{NGC 3628 X-2 (CXOU J112037.3+133429)} 

This source was first detected as ``point source 2'' in a December
1979 \ein IPC image of NGC 3628, as reported by Bregman \& Glassgold
(1982), where it was observed to lie at the eastern end of the edge-on
galaxy disc.  They derived a 0.3 - 2.9 keV luminosity of $2.1 \times
10^{39} \ergsec$ for NGC 3628 X-2, corrected to our assumed a distance
of 7.7 Mpc (this correction is assumed for all the following
luminosities).  However, the reprocessed IPC data contour map
presented by Fabbiano, Kim \& Trinchieri (1992) suggests, in
retrospect, that this flux may be an upper limit due to confusion with
two X-ray sources since resolved to the north of X-2\footnote{These
sources are now identified as the QSO J112041.7+133552 (Veron-Cetty \&
Veron 2000), and a narrow emission-line galaxy RIXOS 259- 7 (Mason et
al. 2000).}.  It was also detected in three separate \ro observations,
once with the PSPC and twice with the HRI, as reported in Dahlem,
Heckman \& Fabbiano (1995) and Dahlem et al. (1996).  In particular,
the November 1991 PSPC observation provides a measurement of its X-ray
luminosity in the 0.1 - 2 keV band of $0.6 \times 10^{39} \ergsec$,
several times lower than the \ein value.  This apparent variability is
supported by the \ro HRI images, which demonstrate that the count rate
of NGC 3628 X-2 drops substantially between December 1991 and May
1994.  NGC 3628 X-2 was also detected in a December 1993 \asca
observation (Yaqoob et al. 1995).  Its \asca SIS spectrum was well-fit
with a simple powerlaw continuum model, with $\Gamma \sim 2.4$ and an
absorption column of \nh $\sim 7 \times 10^{21} \atpcm$, well in
excess of the foreground Galactic column.  This model implies
luminosities of 0.7 and $1.7 \times 10^{39} \ergsec$ in the 0.5 -- 2
and 2 -- 10 keV bands respectively, consistent with the \ro PSPC
measurement in the soft X-ray regime, though this must again be
regarded as an upper limit due to the likelihood of confusion with
other sources in the wide \asca beam.  Yaqoob et al. (1995) comment
that the X-ray spectrum of NGC 3628 X-2 is consistent with either a
low-mass X-ray binary or a very young supernova remnant, noting that
if it is an accreting source then it is substantially super-Eddington
for a 1 M$_{\odot}$ object.  Finally, NGC 3628 X-2 was not covered in
the previous \chan ACIS-S observation of NGC 3628 discussed by
Strickland et al. (2001), as their observation was aligned along the
minor axis of the galaxy.

\subsection{NGC 4136 X-1 (CXOU J120922.6+295551)}

The first and, previous to this work, only detection of this ULX was
made by Lira, Lawrence \& Johnson (2000), who obtained a luminosity of
$2.5 \times 10^{39} \ergsec$ (corrected to a distance of 9.7 Mpc) from
a \ro HRI observation of NGC 4136.  They note that its position is
coincident with a diffuse blue optical counterpart, which may be
emission knots in the spiral arms of the galaxy.

\subsection{NGC 4559 X-1 (CXOU J123551.7+275604) \& X-4 (CXOU
J123558.6+275742)}

NGC 4559 X-1 was first detected in the \ro all-sky survey bright
source catalogue as 1RXS J123551.6+275555 (Voges et al. 1999).  It was
identified with the outer edge of the galactic disc of NGC 4559 by
Vogler, Pietsch \& Bertoldi (1997), who present a detailed analysis of
the X-ray properties of this source based on a \ro PSPC pointed
observation.  In their work they refer to this source as NGC 4559 X-7.
The PSPC profile of the source is point-like, and no significant X-ray
variability is detected from it during the observation.  Its spectrum
is well-fit by either a powerlaw continuum ($\Gamma \sim 3$) or
thermal bremsstrahlung model ($kT \sim 0.8$ keV), both showing
absorption \nh $\sim 1 - 2 \times 10^{21} \atpcm$, well in excess of
the foreground galactic column.  The observed (0.1 - 2.4 keV)
luminosity of $6 \times 10^{39} \ergsec$ converts to an impressive
intrinsic luminosity of $1.5 \times 10^{40} \ergsec$ using the thermal
bremsstrahlung model.  The PSPC position is coincident with a group of
emission knots in a faint outer spiral arm of the galaxy, which Vogler
et al. identify as likely \hii regions.  It is, however, too bright to
be the integrated X-ray emission of ``ordinary'' X-ray binaries and
supernova remnants associated with the \hii region.  They conject that
this source is most likely a supernova remnant buried in a dense cloud
of interstellar material, though they do not rule out either a
black-hole accreting binary system, or mini-AGN from a merging dwarf
galaxy, as alternative identifications.  This source was also
detected, with a similar high X-ray luminosity, in a \ro HRI
observation (RW2000).

Unlike NGC 4559 X-1, NGC 4559 X-4 was not detected in the \ro all-sky
survey bright source catalogue.  However, in the \ro PSPC observation
reported by Vogler et al. (who list this ULX as NGC 4559 X-10) it was
brighter than X-1.  It is located near, but not at, the centre of the
galaxy.  It appears as an extended source in the PSPC observation,
with about 20\% of its flux emanating from between 20 and 50
arcseconds from the source centroid, which may be a diffuse component,
or confusion with other point sources (with our \chan data suggesting
the latter).  The X-ray spectrum is well-fit by either a powerlaw
continuum, or thermal bremsstrahlung emission, albeit spectrally
harder in each case than for X-1 ($\Gamma \sim 2.0$, or $kT \sim 2.1$
keV, respectively), with a measured absorption column of \nh $\sim
10^{21} \atpcm$.  This gave an observed (intrinsic) luminosity of $7
(12) \times 10^{39} \ergsec$.  Vogler et al. suggest that X-10 is the
superposition of several point sources, though they cannot rule out
the contribution of an AGN towards the observed flux.  This source is
again also detected in the \ro HRI survey of RW2000, offset from the
nucleus of NGC 4559 by 13 arcseconds.

\subsection{NGC 5204 X-1 (CXOU J132938.6+582506)}

This ULX was first detected in an \ein IPC observation, with a 0.2 --
4 keV X-ray luminosity of $\sim 2.6 \times 10^{39} \ergsec$ (Fabbiano, 
Kim \& Trinchieri 1992).  It also appeared in the \ro all-sky survey
bright source catalogue with a count rate of $7.5 \times 10^{-2}
\ctsec$.  Later \ro HRI pointed observations revealed NGC 5204 X-1 to
lie $\sim 20$ arcseconds to the east of the nucleus of NGC 5204
(Colbert \& Mushotzky 1999; RW2000; Lira, Lawrence \& Johnson 2000).
This ULX was revealed to have a point-like optical counterpart with a
blue continuum spectrum in a William Herschel Telescope integral field
observation (Roberts et al. 2001), suggestive of the presence of
several O-stars.  A more detailed study using archival \hst data
resolved the ULX into three possible counterparts, all of which are
consistent with young ($< 10$ Myr old) compact stellar clusters in NGC
5204, suggesting that this ULX may be a high-mass X-ray binary system
(Goad et al. 2002).

\end{document}